\documentclass[11pt,a4paper,pdflatex]{article}
\usepackage{jheppub}
\usepackage{physics,bm}
\usepackage{amsmath,amsfonts,amssymb}
\usepackage{autobreak}
\usepackage{hyperref,comment}
\usepackage{color}
\usepackage{graphicx}
\hypersetup{colorlinks=true, citecolor=blue, urlcolor=blue, linkcolor=blue}
\usepackage{natbib}

\newcommand{\Eac}{E_{\text{vanish}}}
\newcommand{\Enear}{E_{\text{nrst}}}

\title{Stark effect and dissociation of mesons in holographic conductor}

\author[a]{Shuta~Ishigaki}
\author[b]{Shunichiro~Kinoshita}
\author[c]{Masataka~Matsumoto}
\affiliation[a]{Department of Physics, Shanghai University,\\
	99 Shangda road, Shanghai 200444, China}
\affiliation[b]{Department of Physics, College of Humanities and Sciences, Nihon University,\\
	Tokyo 156-8550, Japan}
\affiliation[c]{Wilczek Quantum Center, School of Physics and Astronomy, Shanghai Jiao Tong University,\\
	Shanghai 200240, China}

\emailAdd{shutaishigaki@shu.edu.cn, kinoshita.shunichiro@nihon-u.ac.jp, masataka@sjtu.edu.cn}
\abstract{
	We study the meson spectrum of the ${\cal{N}}=4$ supersymmetric Yang-Mills theory with ${\cal{N}}=2$ fundamental hypermultiplets for a finite electric field by using the D3/D7 model.
	The spectrum for scalar and vector mesons is computed by analyzing the (quasi-)normal modes for the fluctuations of the D7-brane embedding and gauge fields.
	In the presence of an electric field, two different phases in the background are realized: the meson and dissociation phases.
	In this paper, we analyze the meson spectrum of scalar and vector mesons for all ranges of the electric field and explore the effect of the electric field on the meson spectrum, that is, the Stark effect.
	In the meson spectrum, we observe the avoided crossing between different levels due to the coupling of fluctuations via the electric field.
}

\begin{document}
	\maketitle

	\section{Introduction}
	For decades, the AdS/CFT correspondence (or holography) \cite{Maldacena:1998,Gubser:1998,Witten:1998} has been employed as one of the powerful tools for non-perturbative approaches to the strongly correlated gauge theory.
	It conjectures the equivalence between a classical gravity theory and quantum gauge theory with a strong correlation.
	In particular, the correspondence arises from the two different perspectives of a stack of $N_{c}$ D3-branes: IIB string theory on AdS$_{5}\times S^{5}$ and ${\cal{N}}=4$ supersymmetric Yang-Mills (SYM) theory.
	
	As discussed in \cite{Karch:2002sh}, the flavor degrees of freedom can be imposed by introducing $N_{f}$ D7-branes with a stack of $N_{c}$ D3-branes in ten-dimensional Minkowski spacetime. 
	Taking the large-$N_{c}$ limit ($N_{c} \gg 1)$ and large 't~Hooft limit ($g_{\rm YM}^{2}N_{c}\gg1$), the D3-branes are replaced by the near-horizon AdS$_{5}\times S^{5}$ geometry. Besides, if we consider the probe limit $N_{c}\gg N_{f}$, the D7-branes can be treated as probes in this geometry.
	The corresponding description is the ${\cal{N}} =4 $ $SU(N_{c})$ SYM theory with the ${\cal{N}} =2 $ $N_{f}$ hypermultiplets fields.
	In this paper, we consider $N_{f}=1$ for simplicity. 
	
	The spectrum of mesons, namely quark-antiquark bound states, in the ${\cal{N}}=2$ theory has been intensively explored by considering the fluctuations of the D7-brane~\cite{Kruczenski:2003be} (see review \cite{Erdmenger:2007cm} and references therein). 
	A natural question to ask is how the meson spectrum will change in the presence of an external electric/magnetic field.
	The effect of an external magnetic field on the meson spectrum, that is, the Zeeman effect, was studied in \cite{Filev:2007gb,Albash:2007bk}.
	On the other hand, the fate of the meson spectrum in the presence of an external electric field was partially studied in \cite{Erdmenger:2007bn} but is still not fully understood.
	A qualitative difference from applying magnetic fields is that the system (with no charge density) can be driven into a non-equilibrium steady state with a constant electric current by applying large electric fields.
	Then, the mesons are no longer stable and are dissociated into independent quarks and anti-quarks with finite lifetime, which is called meson melting \cite{Albash:2007bq}. (Also, the meson melting at high temperature was discussed in~\cite{Hoyos-Badajoz:2006dzi}.)
	In other words, the corresponding fluctuations on the D$7$-brane are no longer described by normal modes but quasi-normal modes (QNMs), whose eigenfrequencies are complex.
	
	Moreover, the electric field makes it difficult to identify the origin of each meson because the external field induces the coupling between mesons.
	In this paper, we focus on the bosonic modes at lower levels on the D7-brane corresponding to the light mesons: one scalar, one pseudo-scalar, and one vector meson with three degrees of freedom corresponding to three spatial components.
	The spectra of these five modes are degenerate in vacuum at zero electric field.
	However, the scalar and one component of the vector, along the electric field, are coupled via the applied electric field.
	Since the electric field causes both the dissociation of the mesons and the coupling between the scalar and vector mesons, the meson spectrum for finite electric field becomes quite complicated.

%	In our previous study, we discuss the dynamical instability of the meson dissociation phase by focusing on the purely imaginary modes \cite{Ishigaki:2021vyv}.
%	However, there are other characteristic modes which appear close to the real axis even near the dissociation point (figure 7 in \cite{Ishigaki:2021vyv} or figure XX in section \ref{sec:Results} of this paper).
%	These modes have not been identified but in this paper we find that they correspond to the lower excitations of mesons.

	Motivated by this, we analyze the meson spectrum, including the decoupling mesons, for all the ranges from the zero electric field to the strong electric field limit (corresponding to the zero quark-mass limit) at zero temperature.
	The fluctuations on the D$7$-brane are decoupled in the massless limit as well as at zero electric field.
 	We will identify the discrete mass spectrum based on each meson excitation decoupled at zero electric field or the massless limit.
%	Motivated by this, we analyze the meson spectrum, including the decoupling mesons, for all range of electric field at zero temperature, in order to identify each meson in the system and investigate the effect of the electric field on them.
	Turning on the electric field lifts the degeneracy and splits each spectrum, known as the Stark effect.
	In this paper, we study each meson spectrum from the meson phase to the limit of large electric field in the dissociation phase.
	 Note that the spectrum and Stark effect for one of the decoupled modes, pseudo-scalar, in the meson phase was studied in \cite{Erdmenger:2007bn}.

	Here, we mention the characteristic modes observed, which appear close to the real axis even near the dissociation point in the previous paper (figure 7 in \cite{Ishigaki:2021vyv} or figure \ref{fig:qnm} in section \ref{sec:Results} of this paper).
	In this paper, we identify them as the lower excitations of mesons as denoted in figure \ref{fig:qnm}.
	
	Before closing this section, we would like to comment on the connection to condensed matter physics.  In the D3/D7 model, strictly speaking, the above mentioned electric/magnetic fields couple to the global U(1) baryon charge of quarks. However, those can be naturally considered as the external fields coupled to charged careers in condensed matter systems. In this sense, the meson melting corresponds to the dielectric breakdown of the insulator and the system is in the steady state with the constant electric current. The meson excitations can be interpreted as the excitons of electrons and holes.
	
	This paper is organized as follows.
	In section \ref{sec:Setup}, we briefly review the D3/D7 model with an electric field.
	We discuss the background solutions of the insulating and conducting phases, corresponding to the meson and dissociation phases, respectively.
	 In addition, we uncover the discrete self-similar behavior of the brane embeddings near the critical solution with the electric field.
	In section \ref{sec:Perturbation}, we consider the fluctuations of the background solutions and explain the method to compute the meson spectrum.
	In section \ref{sec:Results}, we show the meson spectrum and discuss its characteristic properties such as an avoided crossing.
	We also present analytical approaches to the meson spectrum.
	Section \ref{sec:Conclusion} is devoted to conclusion and discussion.
	
	\section{Background} \label{sec:Setup}
	In this section, we briefly review the setup of the D3/D7 model.
	For our purpose, we apply an external electric field on a probe D7-brane and consider the AdS\(_5 \times S^5\) spacetime as a background geometry.
	In this situation, the dual system corresponds to %the many-body charged particles system at zero temperature.
	fundamental matter including quark/antiquark degrees of freedom in the electric field at zero temperature.

	\subsection{D3/D7 model with an external electric field}
	The bulk metric of the $10$-dimensional AdS\(_5 \times S^5\) spacetime is given by%
	\begin{equation}
		\dd s^{2}_{10}= \frac{1}{u^{2}}\left(-\dd t^{2}+\dd\vec{x}^{2}+\dd u^{2}\right)+\dd\Omega_{5}^{2},
		\label{eq:metric}
	\end{equation}
	where $(t,\vec{x})=(t,x,y,z)$ represent the coordinates of the dual gauge theory in the (3+1)-dimensional spacetime and $u$ denotes the radial coordinate of the AdS direction. 
	Here, we set each radius of AdS$_{5}$ and $S^{5}$ parts to 1 for simplicity.
	The AdS boundary is located at $u=0$ in this coordinate.
	%The bulk geometry corresponds to the system at zero temperature.
	The metric of the $S^{5}$ part is given by
	\begin{equation}
		\dd\Omega_{5}^{2}=\dd\theta^{2}+\sin^{2}\theta \dd\psi^{2}+\cos^{2}\theta \dd\Omega_{3}^{2},
	\end{equation}
	where $\dd\Omega_{3}^{2}$ denotes the line element of the $S^{3}$ part.
	We note that this bulk geometry corresponds to the $(3+1)$-dimensional dual field theory at zero temperature.

	%In this study, we introduce a single D7-brane ($N_{f}=1\ll N_{c}$), which fills the AdS$_{5}$ part and wraps the $S^{3}$ part of the $S^{5}$. The configuration of the D7-brane is determined by the embedding functions of $\theta$ and $\psi$.
	
	The action for the probe D7-brane is given by the Dirac-Born-Infeld~(DBI) action:
	\begin{equation}
		S_{\text{D7}}=-T_{\text{D7}} \int \dd^{8}\xi \sqrt{-\det \left(g_{ab}+(2\pi\alpha')F_{ab} \right)}
		+ S_{\text{WZ}},
	\end{equation}
	where \(S_{\text{WZ}}\) denotes the Wess-Zumino term.
	In our study, we can ignore the Wess-Zumino term without loss of generality.
	 (See Appendix \ref{appendix:WZ_term} for details.)
	%Without loss of generality, we ignore the Wess-Zumino term in this study.
	The D7-brane tension is given by 
	$T_{\text D7}=(2\pi)^{-7}(\alpha')^{-4}g_{\text s}^{-1}$ and
	%$T_{\text D7}$ is the tension of the D7-brane given by $T_{\text
		%D7}=(2\pi)^{-7}(\alpha')^{-4}g_{\text s}^{-1}$. 
	$g_{ab}$  is the induced metric defined by
	\begin{equation}
		g_{ab} = \frac{\partial X^{\mu}}{\partial \xi^{a}} \frac{\partial X^{\nu}}{\partial \xi^{b}}G_{\mu\nu},
	\end{equation}
	where $\xi^{a}$ denotes the worldvolume coordinate on the D7-brane with
	$a,b=0,\ldots,7$ and $X^{\mu}$ denotes the target space coordinate with
	$\mu,\nu=0,\ldots,9$. $G_{\mu\nu}$ represents the background metric given by
	eq.~(\ref{eq:metric}). The field strength of the U(1) worldvolume gauge field on the
	D7-brane is given by $2\pi\alpha' F_{ab} \equiv \partial_{a}A_{b}-\partial_{b}A_{a}$.
	
	In this study, we introduce a single D7-brane ($N_{f}=1\ll N_{c}$), which fills the AdS$_{5}$ part and wraps the $S^{3}$ part of the $S^{5}$. The configuration of the D7-brane is determined by the embedding functions of $\theta$ and $\psi$.
	We consider the ansatz for the embedding functions and the worldvolume gauge fields in this study as %is given by
	\begin{equation}
		\theta=\theta(u), \hspace{1em} \psi =\text{const.}, \hspace{1em} A_{x}=-Et + h(u).
		 \label{eq:ansatz_background}
	\end{equation}
	Note that we can take $\psi = 0$ without loss of generality,
	 because the bulk spacetime is axisymetric with respect to $\psi$.
	Then, the induced metric and the field strength are respectively given by 
	\begin{equation}
		g_{ab}\dd\xi^a\dd\xi^b = \frac{1}{u^{2}}
		\left(-\dd t^{2} + \dd\vec{x}^{2}\right)
		%  + \frac{1+u^2f(u)\theta'(u)^2}{u^2f(u)}\dd u^{2}
		+ \left(\frac{1}{u^2} + \theta'(u)^2\right)\dd u^{2}
		+ \cos^2\theta(u) \dd\Omega_{3}^{2} ,
		\label{eq:induced_metric}
	\end{equation}
	and 
	\begin{equation}
		2\pi\alpha' F_{ab}\dd\xi^a\wedge\dd\xi^b = - \frac{E}{2}\dd t\wedge \dd x
		- \frac{h'(u)}{2} \dd x \wedge \dd u ,
		\label{eq:Fab_components}
	\end{equation}
	where the prime denotes the derivative with respect to $u$.
	The DBI action is written as
	\begin{equation}
		\begin{aligned}
			S_{\text D7} &= {\cal N}\int \dd t \dd^{3}x \dd u \, \mathcal{L} , \\
			\mathcal{L} &\equiv - \cos^{3}\theta(u) 
			g_{xx} \sqrt{- g_{tt}g_{xx}g_{uu}- g_{uu}E^{2} - g_{tt} h'(u)^{2}},
		\end{aligned}
		\label{eq:L_DBI}
	\end{equation}
	where the prefactor of the action is given by
	\begin{equation}
		{\cal N}= T_{\text D7} (2\pi^{2}) =\frac{N_{c}\lambda}{(2\pi)^{4}} ,
	\end{equation}
	with the relation $4\pi g_{s} N_{c} \alpha'^{2}=1$ and 't Hooft coupling
	$\lambda = g_{\text YM}^2 N_{c} = 4\pi g_s N_{c}$. 
	%\begin{equation}
	%	S_{\text D7} = -{\cal N}\int \dd t \dd^{3}x \dd u  \, \cos^{3}\theta g_{xx} \sqrt{\abs{g_{tt}}g_{xx}g_{uu}- g_{uu}E^{2} + \abs{g_{tt}} h'^{2}},
	%\end{equation}
	%where the prime denotes the derivative of $u$. Note that $g_{uu}= 1/u^{2} + \theta'(u)^{2}$. Hereafter, we assume $2\pi\alpha' =1$ for simplicity. The prefactor of the action is given by
	%\begin{equation}
	%	{\cal N}= T_{\text D7} (2\pi^{2}) =\frac{N_{c}}{(2\pi)^{2}}.
	%\end{equation}
	%with the relation $4\pi g_{s} N_{c} \alpha'^{2}$=1.
	Near the AdS boundary ($u=0$), the fields can be expanded as follows:
	\begin{equation}
		\begin{aligned}
			\sin\theta(u) &= m u + c u^3 + \cdots,\\
			A_x(t,u) & = - E t + h(u) = - E t + \frac{J}{2} u^2 + \cdots,
			%	\theta(u) &= m_q u - c u^3 + \cdots,\\
			%	A_x(u) & = - E t + \frac{J}{\mathcal{N}} u^2 + \cdots,
		\end{aligned}
		\label{eq:asymptotic}
	\end{equation}
	where, according to the AdS/CFT dictionary, the coefficients $m$ and $c$ are related to quark mass and quark
	condensate~\cite{Mateos:2007vn}, and $E$ and $J$ are related to electric field and
	electric current density~\cite{Karch:2007pd} in the dual field theory: 
	\begin{equation}
		m_q = \frac{\lambda^{1/2}}{2\pi} m ,\quad
		\left<\bar{q}q \right> = 2 \frac{N_c}{(2\pi)^3} \lambda^{1/2} c ,
	\end{equation}
	and 
	\begin{equation}
		\mathcal{E}_x = \frac{\lambda^{1/2}}{2\pi} E ,\quad
		\left<\bar{q} \gamma^x q \right> = \frac{N_c}{(2\pi)^3}\lambda^{1/2} J .
	\end{equation}
	Hereinafter, for simplicity, we will refer to geometrical parameters
	$m$, $c$, $E$, and $J$ as quark mass, quark condensate, electric field, and
	electric current, respectively.

	The D$7$-brane action does not explicitly contain $h(u)$ but
	depend only on $h'(u)$, so that the following quantity 
	\begin{equation}
		\frac{\partial\mathcal{L}}{\partial h'} = 
		\frac{\cos^3\theta g_{tt}g_{xx} h'}
		{\sqrt{(- g_{tt}g_{xx} - E^{2})g_{uu} - g_{tt} h'^{2}}} 
		\label{eq:conserved_quantity}
	\end{equation}
	is conserved with respect to $u$-derivative.
	Substituting the asymptotic behavior (\ref{eq:asymptotic})
	near the AdS boundary, we find that this conserved quantity coincides
	with the electric current density as 
	$J = -\lim_{u\to 0} \partial\mathcal{L}/\partial h'$.
	On the other hand, if there exists the locus $u=u_* \equiv 1/\sqrt{E}$ 
	such that $-g_{tt}g_{xx} - E^2 = 0$, 
	we can obtain 
	\begin{equation}
		\left.\frac{\partial\mathcal{L}}{\partial h'}\right|_{u=u_*}
		= - \cos^3\theta(u_*) \sqrt{-g_{tt}}g_{xx} = -J 
		\label{eq:current_density}
	\end{equation}
	in a straightforward way by evaluating it there.
	Thus, the electric current density $J$ should be determined locally at $u=u_*$.
	This is because satisfying $\partial^2\mathcal{L}/\partial h'^2 = 0$ at $u=u_*$ makes the equations of motion for $h(u)$ locally degenerate.
	It implies that such a locus becomes a characteristic surface for the equations of motion.
	%In fact, the locus $u=u_*$ is referred to as the `` effective horizon'' 
%	on the worldvolume of the D$7$-brane because a causal boundary is given by an effective metric on the
%	worldvolume, which governs dynamics of brane.%
%	\footnote{This fact plays a significant role in studying the time evolution
%		of brane dynamics. For example, see ref.~\cite{Hashimoto:2014yza}.}
	In fact, the locus $u=u_*$ is a causal boundary given by an effective metric on the worldvolume and is referred to as the `` effective horizon'' on the worldvolume of the D$7$-brane.
	It plays a significant role in time evolutions of non-linear dynamics for the brane~\cite{Hashimoto:2014yza}.
	For the DBI action, the effective metric is the so-called open-string
	metric defined by 
	$\gamma_{ab} \equiv g_{ab} + (2\pi\alpha')^2 F_{ac}F_{bd}g^{cd}$
	~\cite{Kim:2011qh,Seiberg:1999vs}:
	\begin{equation}
		\begin{split}
			\gamma_{ab}\dd\xi^a\dd\xi^b 
			&= \frac{g_{tt}g_{xx} + E^2}{g_{xx}} \dd t^2
			- 2\frac{Eh'}{g_{xx}} \dd t \dd u + 
			\frac{g_{xx}g_{uu} + h'^2}{g_{xx}} \dd u^2 \\
			&\quad + \left(g_{xx} + \frac{E^2}{g_{tt}} + \frac{h'^2}{g_{uu}}\right) \dd x^2
			+ \frac{\dd y^2 + \dd z^2}{u^2} + \cos^2\theta(u) \dd\Omega_3^2 .
		\end{split}
		\label{eq:effmetric}
	\end{equation}
	This shows that the locus $u=u_*$ is the Killing horizon with respect
	to time coordinate $t$ on the AdS boundary, which corresponds to the time in
	the dual field theory.
	We also define $\gamma^{ab}$ by $\gamma^{ac}\gamma_{cb} = \delta^{a}{}_{b}$ for later use.
	
	\subsection{D7-brane embeddings}
	We can determine the configuration of the D$7$-brane by solving the equations of motion for $\theta(u)$ and $h(u)$, which are derived from the action (\ref{eq:L_DBI}).
	%To obtain the configuration of the D$7$-brane, we should solve
	%the equations of motion for $\theta(u)$ and $h(u)$, which are given by
	%nonlinear second-order ordinary differential equations.
	There are three types of solutions depending on
	the bulk behavior: the so-called {\em Minkowski}, 
	{\em black hole}, and {\em critical embeddings}~\cite{Mateos:2006nu,Mateos:2007vn,Frolov:2006tc}.
	The solutions can be classified according to whether a horizon emerges on the worldvolume of the
	D$7$-brane.
	
	First, for the Minkowski embeddings, the solutions of the brane embedding
	function $\theta(u)$ can smoothly reach $\theta(u_0)=\pi/2$ ($u_0<u_*$), 
	at which the size of
	$S^3$ wrapped by the D$7$-brane shrinks to zero, without encountering any horizon on
	the brane.
	In this phase, fluctuations on the D$7$-brane cannot be dissipated, meaning that they are generally described by normal modes with real
	frequencies, which correspond to stable excitations of meson, i.e.,
	quark/antiquark bound states in the dual field theory~\cite{Kruczenski:2003be}.
	In addition, we can see that the electric current density $J$
	vanishes and the solution of the gauge field becomes trivial,
	$h(u)=0$, even under external electric fields.
	Thus, we consider the Minkowski embedding phase as the insulating
	state in the dual field theory.
	
	Second, for the black hole embeddings, the effective horizon emerges at 
	$\theta(u_*)<\pi/2$.
	In this phase, fluctuations on the brane can eventually dissipate into
	the effective horizon, described by QNMs with
	complex frequencies, implying that the excitations of meson become unstable, and then
	quark/antiquark bound states will dissociate in the dual field
	theory.%
	\footnote{QNM analysis in the meson melting at
	finite temperature was discussed in ref.~\cite{Hoyos-Badajoz:2006dzi}.}
	Under external electric fields, the current density $J$ becomes finite
	as previously mentioned in eq.~(\ref{eq:current_density}).
	Hence, we can consider the black hole embedding phase as the conducting
	state in the dual field theory. 
	In particular, because Joule heating occurs by the constant
	electric field and current, the system becomes nonequilibrium
	steady state.
	
	Finally, the critical embedding is a critical solution between the
	Minkowski and black hole embeddings. 
	The embedding function $\theta(u)$ reaches $\theta=\pi/2$ at $u=u_*$,
	and the D$7$-brane has a conical configuration with no current density, $J=0$.
	(For example, see ref.~\cite{Hashimoto:2015wpa}.)
	
	Depending on which type of solution we wish to obtain, we should impose the appropriate boundary condition at $u=u_{0}$ or $u=u_{*}$. 
	%At the AdS boundary~$u=0$, we impose the value of \(m_{q}\), namely~\( \left. u\theta(u) \right|_{u\rightarrow 0}=m_{q}\).
	Near the AdS boundary $u=0$, we can read the observables in the dual field
	theory according to eq.~(\ref{eq:asymptotic}).
	
\begin{figure}[tbp]
		\centering
		\includegraphics[width=7cm]{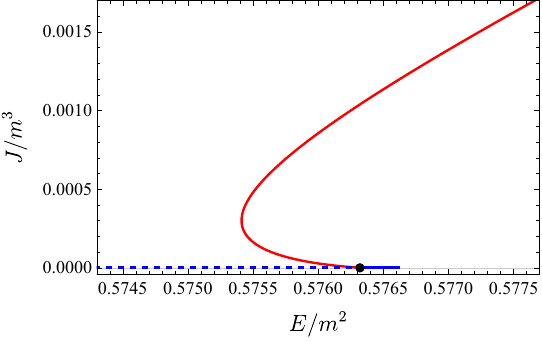}
		\includegraphics[width=7cm]{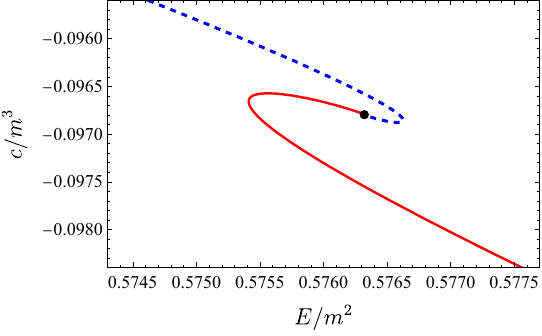}
		\caption{The $E$-$J$ (left) and $E$-$c$ (right) curves at zero temperature in the vicinity of the critical value. The red
	 solid and blue dashed lines denote the insulating and conducting states, respectively. The critical value (black dot) of $E/m^{2}$ is given by $E_{c}/m^{2}\approx0.57632$.}
		\label{fig:EJplot}
	\end{figure}

	Near the critical embedding, we find a transition between the Minkowski and black hole embedding depending on the electric field.
	The transition between two different embeddings in the bulk can be interpreted as that between the insulating and conducting state in the dual field theory, that is, dielectric breakdown by external electric fields.
	In terms of mesons, i.e., bound states of quark and antiquark, the transition corresponds to the dissociation of the mesons.
	In figure \ref{fig:EJplot}, we show the $E$-$J$ and $E$-$c$ curves at zero
	temperature~\cite{Hashimoto:2014yza}. 
	The current density $J$ and the quark condensate $c$ are multi-valued functions of the electric field $E$ at zero temperature, as well as at finite temperature \cite{Nakamura:2010zd}.
	%As well-known in the D3/D7 model at finite
	%temperature, the current density is
	%a multi-valued function of the electric field \cite{Nakamura:2010zd}.
	The critical value of $E/m^{2}$ is given by $E_{c}/m^{2}\approx0.57632$, at which the D$7$-brane becomes the critical embedding. 
	The $E$-$J$ and $E$-$c$ curves at finite temperature $T$ and the phase diagram for $(\pi T/m, E/m^{2})$ are shown in \cite{Ishigaki:2021vyv}.
	
	%\if0
	\subsection{Critical behavior} \label{sec:critical}
	
	As can be seen in figure~\ref{fig:EJplot}, the $E$-$J$ and $E$-$c$
	curves turn over and the current density $J$ and the quark
	condensate $c$ are multivalued for a given electric field $E$.
	Indeed, near the critical solution there coexist multiple
	solutions not only of the black hole embeddings but of the
	Minkowski embeddings for the same parameter.
	Such behavior repeatedly appears as the parameter approaches to
	the critical value. 
	We will discuss critical behavior around the critical embedding solution by following refs.~\cite{Frolov:2006tc,Mateos:2006nu,Mateos:2007vn}.

	We introduce the following bulk coordinates:
	\begin{equation}
	 \rho = u^{-1} \cos\theta ,\quad W = u^{-1} \sin\theta ,
	 \label{eq:cartesian_bulk_coordinate}
	\end{equation}
	and then rewrite the bulk metric (\ref{eq:metric}) as 
	\begin{equation}
	 \dd s^{2}_{10}= (\rho^2 + W^2)\left(-\dd
					t^{2}+\dd\vec{x}^{2}\right)+
	 \frac{1}{\rho^2 + W^2} 
	 \left(
	  \dd\rho^2 + \dd W^2 + \rho^2\dd\Omega_{3}^{2} + W^2 \dd\psi^2
	 \right) .
	 \label{eq:cartesian_bulk_metric}
	\end{equation}
	In this coordinate system, 
	the embedding functions of a probe D$7$-brane are given by 
	\begin{equation}
	 W = W(\rho(u)) = \frac{\sin\theta(u)}{u} , \quad \rho(u) = \frac{\cos\theta(u)}{u} .
	  % W = W(\rho) ,\quad \psi = \text{const.}
	\end{equation}
 	The induced metric on the D$7$-brane is 
	\begin{equation}
	 g_{ab}\dd \xi^a \dd \xi^b = [\rho^2 + W(\rho)^2] (- \dd t^2 + \dd \vec{x}_3^2)
	  + \frac{1}{\rho^2 + W(\rho)^2}[(1+(dW/d\rho)^2)\dd\rho^2 + \rho^2
	  \dd\Omega_3^2] .
	\end{equation}
 %We suppose that the worldvolume gauge field is 
 %$A_a dy^a = [- Et + a(\rho)]dx$ and the embedding function is
 %$w(\rho) = \sqrt{E} + \xi(\rho)$.
 In order to analyze near-critical solutions, we rewrite the embedding function as $W(\rho) = u_*^{-1} + f (\rho) = \sqrt{E} + f(\rho)$ 
and suppose that $f(\rho)=\epsilon \tilde{f}(\rho/\epsilon)$
 and $h(\rho)=\epsilon^{3/2}\tilde{h}(\rho/\epsilon)$.
 %Assuming $E \gg \sqrt{E}f \gg \rho^2$, we can rewrite a Lagrangian density of the DBI action (\ref{eq:L_DBI}) near $\rho=0$ as  
 Assuming $\epsilon /\sqrt{E} \ll 1$ and considering the near region 
$0\le \rho \equiv \epsilon\tilde\rho \lesssim \sqrt{E}$, the Lagrangian density of the DBI action (\ref{eq:L_DBI}) at leading order in $\epsilon$ is given by 
\begin{equation}
 \begin{aligned}
  \mathcal{L}\frac{du}{d\rho} &= - \rho^3\sqrt{1+(dW/d\rho)^2}
  \sqrt{1 - \frac{E^2}{(\rho^2 + W(\rho)^2)^2}
  + \frac{(dh/d\rho)^2}{1+(dW/d\rho)^2}
  } \\
  &\sim \tilde\rho^3 
  \sqrt{\tilde{f}(\tilde\rho) \left[1+\left(\frac{d\tilde{f}}{d\tilde\rho}\right)^2\right]
  + \left(\frac{d\tilde{h}}{d\tilde\rho}\right)^2} .
 \end{aligned}
\end{equation}
This system has a scaling symmetry: $\tilde{f}(\tilde\rho) \to \tilde{f}(\mu\tilde\rho)/\mu$ 
and $\tilde{h}(\tilde\rho) \to \tilde{h}(\mu\tilde\rho)/\mu^{3/2}$.
Once a set of solutions for $\tilde{f}(\tilde\rho)$ and $\tilde{h}(\tilde\rho)$ is obtained, one can generate other solutions by the scaling transformation for any real positive $\mu$. 

First of all, it turns out that the equations of motion given by this effective Lagrangian have exact solutions 
\begin{equation}
 \tilde{f}_\mathrm{c}(\tilde\rho) = \frac{1}{\sqrt{6}}\tilde\rho ,\quad \tilde{h}_\mathrm{c}(\tilde\rho) = 0 ,
\end{equation}
which correspond to the critical embedding solution with conical configuration near $\rho = 0$.
This implies the cone is locally $\mathbf{R}^+ \times S^3$ with the half-angle $\arctan\sqrt{6}$ at the apex, which is determined only by the force balance~\cite{Hashimoto:2015wpa}. 

Then, we consider linear perturbations around the critical exact solution as 
$\tilde{f}(\tilde\rho) = \tilde{f}_\mathrm{c}(\tilde\rho) + \delta f$ and 
$\tilde{h}(\tilde\rho) = \tilde{h}_\mathrm{c}(\tilde\rho) + \delta h$.
The equations of motion for the perturbations are given by 
\begin{equation}
 \frac{d^2}{d\tilde\rho^2} \delta f + \frac{7}{2\tilde\rho} \frac{d}{d\tilde\rho}\delta f
  + \frac{7}{2\tilde\rho^2}\delta f =0,
  \quad
  \frac{d^2}{d\tilde\rho^2} \delta h + \frac{5}{2\tilde\rho}\frac{d}{d\tilde\rho}\delta h = 0 .
\end{equation}
In near-critical cases, the general solutions for large $\tilde\rho$ behave as%
\footnote{
The solution for the perturbations is of the form 
$\delta f = \tilde{\rho}^{\nu_\pm}$ with complex exponents 
$\nu_\pm = (-5 \pm i \sqrt{31})/4$.
It is worth noting that this exponent near the effective horizon is different from one near the bulk spacetime horizon in the brane-black hole systems, which depends only on the dimension of the internal sphere wrapped by the brane~\cite{Frolov:2006tc,Mateos:2006nu,Mateos:2007vn}.
This implies that the critical behavior differs between transitions due to external electric fields and due to thermal effects in a finite temperature.
}
\begin{equation}
 \begin{aligned}
  \tilde{f}(\tilde\rho) &= \frac{1}{\sqrt{6}}\tilde\rho + C_1 \tilde\rho^{-5/4} \sin
  \left(\frac{\sqrt{31}}{4}\log \tilde\rho\right)
  +  C_2 \tilde\rho^{-5/4} \cos
  \left(\frac{\sqrt{31}}{4}\log \tilde\rho\right) ,\\
  \tilde{h}(\tilde\rho) &= C_3 + C_4 \tilde\rho^{-3/2} .
 \end{aligned}
\end{equation}
The relation between the coefficients $C_n$ should be determined by the boundary conditions near $\tilde\rho \ll 1$, while they are boundary conditions to match %the far-region solutions for $\rho \gtrsim \sqrt{E}$. 
full non-linear solutions extending to the far region $\rho \gtrsim \sqrt{E}$.
The scaling symmetry makes the coefficients transform as 
\begin{equation}
 \begin{pmatrix}
  C_1 \\ C_2
 \end{pmatrix}
 \to \frac{1}{\mu^{9/4}}
 \begin{pmatrix}
  \cos\left(\frac{\sqrt{31}}{4}\log \mu\right) &
  - \sin\left(\frac{\sqrt{31}}{4}\log \mu\right) \\
  \sin\left(\frac{\sqrt{31}}{4}\log \mu\right) &
  \cos\left(\frac{\sqrt{31}}{4}\log \mu\right)
 \end{pmatrix}
  \begin{pmatrix}
  C_1 \\ C_2
 \end{pmatrix}
 ,
\end{equation}
\begin{equation}
 \begin{pmatrix}
  C_3 \\ C_4
 \end{pmatrix}
 \to 
 \begin{pmatrix}
  1/\mu^{3/2} & 0 \\
  0 & 1/\mu^{3}
 \end{pmatrix}
  \begin{pmatrix}
  C_3 \\ C_4
 \end{pmatrix}
 .
\end{equation}
This implies that the brane embeddings show discrete self-similarity with a period given by $\mu = \mu_{2\pi} \equiv \exp(8\pi / \sqrt{31})$ when approaching the critical value.
It yields the oscillatory behavior of the $E$-$J$ and $E$-$c$ curves in the vicinity of the critical electric field.

\begin{figure}[tbp]
	\centering
	\includegraphics[width=12cm]{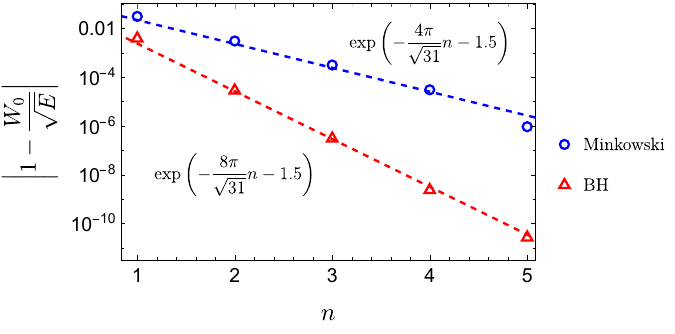}
	\caption{The values of the boundary condition $\abs{1-W_{0}/\sqrt{E}}$ at the $n$-th turning point of the $E$-$J$ and $E$-$c$ curves. The limit $n\to\infty$ corresponds to the critical embedding.
 %The number of turning points $n$ and the corresponding values of the boundary condition $\abs{1-W_{0}/\sqrt{E}}$. 
Each turning point is obtained from the numerical calculations. The dashed lines denote semi-analytical predictions for the period based on the discrete self-similarity, in which the values with $n\to 0$ are determined by the fitting. Note that $W_{0}= \left. W\right|_{\rho=0}$ for the Minkowski embedding and $W_{0}=\left. W \right|_{\rho=\rho_{0}}$ for the black hole (BH) embedding as denoted in (\ref{eq:bdryscale}).}
	\label{fig:scale}
\end{figure}

Now, the boundary condition of the brane embedding function $W(\rho)$ 
%at the center $\rho = 0$ or at the effective horizon $\rho=\rho_0$  
is determined by 
\begin{equation}
  f(\rho) \equiv W(\rho) - \sqrt{E} = f_0 + f_2 \rho^2 + \cdots ,
\end{equation}
at the center $\rho=0$ for the Minkowski embeddings and 
\begin{equation}
  f(\rho) \equiv W(\rho) - \sqrt{E-\rho^2_0} = f_1 (\rho - \rho_0) + \cdots
\end{equation}
at the effective horizon $\rho=\rho_0$ for the BH embeddings.
%\begin{equation}
% \begin{aligned}
%  f(\rho) \equiv W(\rho) - \sqrt{E} = f_0 + f_2 \rho^2 + \cdots ,\\
%  f(\rho) \equiv W(\rho) - \sqrt{E-\rho^2_0} = f_1 (\rho - \rho_0) + \cdots
% \end{aligned}
%\end{equation}
Thus, the scaling transformation rescales the boundary condition for $W(\rho)$ as  
\begin{equation}
 W|_{\rho=0} - \sqrt{E} \sim f_0/\mu \quad (\text{Minkowski}),\quad \sqrt{E} - W|_{\rho=\rho_0} \sim \rho_0^2/\mu^2 \quad (\text{BH}).
 \label{eq:bdryscale}
\end{equation}
%For example, we find the turning points of the $E$-$J$ and $E$-$c$ curves will appear every half-period $\frac{1}{2}\log\mu_{2\pi}$, as shown in figure \ref{fig:scale}.
For example, we find the turning points of the $E$-$J$ and $E$-$c$ curves will appear every half-period $\frac{1}{2}\log\mu_{2\pi}$.
As shown in figure \ref{fig:scale}, on the $n$-th turning point the values of the boundary condition scale as $(W|_{\rho=0} - \sqrt{E})_n \propto 1/\mu_{2\pi}^{n/2}$ and $(\sqrt{E} - W|_{\rho=\rho_0})_n \propto 1/\mu_{2\pi}^n$ for the Minkowski and BH embeddings, respectively.

\section{Linear perturbations}\label{sec:Perturbation}

In this section, we formulate linear perturbations on the probe
D7-brane in order to elucidate the spectra of mesonic excitations.
%For convenience we introduce the following Cartesian-like coordinates in the bulk:
%\begin{equation}
%\rho = u^{-1} \cos\theta,\quad
%W = u^{-1} \sin\theta .
%\end{equation}
%Then, the static brane embedding $W = \bar{W}(u)$ is rewritten as 
%\begin{equation}
%\bar{W}(u) = \frac{\sin\theta(u)}{u} .
%\end{equation}

\subsection{Perturbation fields}

Now, we consider the following ansatz for perturbed fields in terms of the bulk coordinates (\ref{eq:cartesian_bulk_coordinate}):
\begin{equation}
	\begin{gathered}
		W \to \bar{W}(u) + \epsilon w(t, u),\quad
		\psi \to  \epsilon \psi(t, u),\\
		A_{a} \dd \xi^a \to (-E t + h(u)) \dd x + \epsilon a_{\parallel}(t, u)\dd{x} + \epsilon \vec{a}_{\perp}(t,u) \cdot \dd{\vec{x}_{\perp}},
	\end{gathered}
\label{eq:perturbation_ansatz}
\end{equation}
where $\vec{x}_{\perp} = (y,z)$ are perpendicular to the $x$-direction along the background electric field, and $\epsilon$ is a small bookkeeping parameter.
$\bar{W}(u) = \sin\theta(u)/u$ denotes the static brane embedding.
We have assumed that the perturbations do not depend on the spatial coordinates parallel to the AdS boundary and the coordinates on the $S^3$ wrapped by the D$7$-brane. 
It means the corresponding states have no spatial momentum or $\mathrm{R}$-charge.%
\footnote{
	In this paper, we focus on the bosonic degrees of freedom.
	The fermionic degrees of freedom we have ignored correspond to those of the mesino in the boundary theory.
	The fermionic action of the D3/D7 model was given by \cite{Martucci:2005rb}, and the mesino spectra in the vacuum were studied in \cite{Kirsch:2006he} (with some assumptions).
	A more complete discussion for the massive case can be found in \cite{Abt:2019tas}.
	%In a recent paper \cite{Abt:2019tas}, they provided a more complete discussion for the massive case, and studied about generating a light fermionic state by double-trace boundary deformations.
	%In a recent paper \cite{Abt:2019tas}, it was studied that a light fermionic state can be generated by adding a higher dimensional operator, and it was proposed as a generating mechanism of light baryons.
}
In this setup, the perturbation of $A_{t}$ cannot be dynamical, so that we have not listed.%
\footnote{A static perturbation for $A_t$ means adding the charge density to the system.}
The quadratic action with the above ansatz is given by
\begin{equation}
\begin{aligned}
	S^{(2)} =&
	\frac{\mathcal{N}}{2}
	\int \dd[4]{x}\dd{u} \mathcal{L} g^{\perp\perp} \gamma^{\alpha\beta} \partial_{\alpha} \vec{a}_{\perp} \partial_{\beta} \vec{a}_{\perp} \\
	+&\frac{\mathcal{N}}{2}
	\int \dd[4]{x}\dd{u} \mathcal{L} \sin^2\theta(u) \gamma^{\alpha\beta} \partial_{\alpha} \psi \partial_{\beta} \psi \\
	+&\mathcal{N}\int \dd[4]{x}\dd{u} \Big[
		\partial_{\alpha} \Phi^T A^{\alpha\beta} \partial_{\beta} \Phi
  		+ \Phi^T B^{\alpha} \partial_{\alpha} \Phi
  		+ \Phi^T C \Phi
	\Big],
	\label{eq:effective_action}
\end{aligned}
\end{equation}
where $\Phi =( w, a_\parallel )^{T}$, \(\alpha, \beta\) denote the coordinates $(t, u)$, and $A^{\alpha\beta}, B^{\alpha}, C$ are coefficient matrices shown in appendix \ref{appendix:eom_pert} in detail.
The perturbations $\vec{a}_{\perp}$ and $\psi$ are decoupled from the other perturbations so they can be analyzed solely.
By contrast, the perturbations $w$ and $a_\parallel$ are coupled since the coefficient matrices have non-diagonal components. We need to analyze $w$ and $a_\parallel$ together as those as studied in ref.~\cite{Ishigaki:2021vyv}.
In the following, we focus on the coupled sector but we can also consider the decoupled sector in the same way.

Considering the Fourier transformation with respect to \(t\), we write $\Phi$ by
\begin{equation}
	\Phi(t, u)
	=
	\int \frac{\dd{\omega}}{2 \pi} e^{-i \omega t } \tilde{\Phi}_{\omega}(u).
\end{equation}
The quadratic action will take the following form in the Fourier space: 
%The relation to (\ref{eq:effective_action}) is given in Appendix \ref{appendix:eom_pert}.
\begin{equation}
	\begin{aligned}
		S^{(2)} =&\mathcal{N}
		\int_{-\infty}^{\infty}\frac{\dd{\omega}}{2 \pi}
		\int\dd{u}
		\Big[
		(\partial_u \bar{\Phi}_{-\omega}^{T}) \bar{A} \partial_u \bar{\Phi}_{\omega}
		+ \bar{\Phi}_{-\omega}^{T} \bar{B}_{\omega} \partial_u \bar{\Phi}_{\omega}
		+ \bar{\Phi}_{-\omega}^{T} \bar{C}_{\omega} \bar{\Phi}_{\omega}
		\Big] ,
	\end{aligned}
	\label{eq:quadratic_action_momentum}
\end{equation}
where the relation to (\ref{eq:effective_action}) is given in Appendix \ref{appendix:eom_pert}.
We have dropped the infinite volume of $\mathbf{R}^3$, and have redefined the perturbation fields $\bar{\Phi}_{\omega}$ to make the boundary values $\order{1}$.
Since both $w$ and $a_{\parallel}$ have asymptotic expansions starting from $\order{1}$, we do not need to perform this redefinition in our case.
We can simply set $\bar{\Phi}_{\omega} = \tilde{\Phi}_{\omega}$.
We formally write the boundary values of $\bar{\Phi}_{\omega}$ as $\varphi_{\omega}$, that is, $\lim_{u\to0} \bar{\Phi}_{\omega} = \varphi_{\omega}$,
which correspond to the sources of the dual operators.
We also consider the following form of counterterms with a cutoff scale $\epsilon$:%
\footnote{
	The counterterms in the D3/D7 model were given by \cite{Karch:2005ms}
	 and \cite{Karch:2007pd}. 
}
\begin{equation}
	S^{(2)}_\mathrm{ct} = \mathcal{N}\int_{-\infty}^{\infty}\frac{\dd{\omega}}{2 \pi} \varphi^{T}_{-\omega} \bar{C}_{\omega, \epsilon} \varphi_{\omega}.
\end{equation}
The renormalized quadratic action is given by $S^{(2)}_\mathrm{ren} = S^{(2)} + S_\mathrm{ct}^{(2)}$.
$\bar{A}$ is usually a Hermitian matrix, and independent of $\omega$.
The other matrices satisfy $\bar{B}_{-\omega} = \bar{B}_{\omega}^{*}$ and the same for $\bar{C}_{\omega}$ and $\bar{C}_{\omega,\epsilon}$.
Rewriting (\ref{eq:quadratic_action_momentum}) to integrate over only the positive frequency, we obtain
\begin{equation}
	\begin{aligned}
		S^{(2)} = \mathcal{N}
		\int_{0}^{\infty}\frac{\dd{\omega}}{2 \pi}
		\int\dd{u}
		\Big[
		2 (\partial_u \bar{\Phi}_{-\omega}^{T}) \bar{A}^{H} \partial_u \bar{\Phi}_{\omega}
		+ \bar{\Phi}_{-\omega}^{T} \bar{B}_{\omega} \partial_u \bar{\Phi}_{\omega}
		+ (\partial_u \tilde{\Phi}_{-\omega}^{T}) \bar{B}^{\dagger}_{\omega} \bar{\Phi}_{-\omega}
		+ 2 \bar{\Phi}_{-\omega}^{T} \bar{C}^{H}_{\omega} \bar{\Phi}_{\omega}
		\Big],
	\end{aligned}
	\label{eq:quadratic_action_momentum_positive}
\end{equation}
and for the counterterms
\begin{equation}
	S^{(2)}_\mathrm{ct} = \mathcal{N}
	\int_{0}^{\infty}\frac{\dd{\omega}}{2 \pi} 2 \varphi^{T}_{-\omega} \bar{C}_{\omega, \epsilon}^{H} \varphi_{\omega},
\end{equation}
where $M^{H}=(M + M^{\dagger})/2$ denotes a Hermitian part of a matrix.

From (\ref{eq:quadratic_action_momentum_positive}), we obtain the equations of motion in the Fourier space as
\begin{equation}
	\partial_{u} \left(
		2 \bar{A}^{H} \partial_{u} \bar{\Phi}
		+ \bar{B}_{\omega}^{\dagger} \bar{\Phi} \right)
	- \bar{B}_{\omega} \partial_{u} \bar{\Phi}
	- 2 \bar{C}_{\omega}^{H} \bar{\Phi} = 0.
\end{equation}
These equations of motion are given by linear ODEs with the parameter $\omega$, and we can solve them by imposing appropriate boundary conditions for each background solution.
For the BH embeddings, we can determine the physical boundary condition for the perturbations at $u=u_*$ since $\gamma_{\alpha\beta}$ plays a role of the effective metric in the action.
%The regular condition at $u=u_*$ was imposed which is equivalent to the ingoing-wave condition at the effective horizon for the both of the decoupled \cite{Ishigaki:2020coe} and coupled sectors.
%We also impose the regular condition at $u=u_*$ in this study.
We will impose the regular condition at $u=u_*$, which is equivalent to the ingoing-wave condition at the effective horizon for both the coupled \cite{Ishigaki:2021vyv} and decoupled sectors \cite{Mas:2009wf,Ishigaki:2020coe}.
For the Minkowski embeddings, the radius of $S^3$ vanishes at finite $u=u_0 < u_*$: $\cos\theta(u_0) = 0$.
In this case, $u=u_0$ becomes a regular singular point of the linearized equations, and we can impose the physical boundary condition by the regular condition at $u=u_0$. As a result, the physical boundary condition at the IR endpoint of the integral region is always given by the regular condition of the perturbation fields in our formalism.

To find the mass spectra, we need to impose a boundary condition at the other boundary, a vanishing Dirichlet condition: $\varphi_{\omega} = 0$.
Then, the problem becomes solving the two-point boundary value problem with $\omega$.
The eigenvalue $\omega$ can be understood as a mass spectrum.
In this study, we compute such $\omega$ by using the collocation method.
See appendix \ref{appendix:collocation} for the details of the computational method.

\subsection{Green's functions and residues in the boundary theory}
For later use, we define a solution matrix of the perturbation fields.
In refs.~\cite{Amado:2009ts,Kaminski:2009dh}, this formulation was utilized for solving two point boundary value problems by the shooting method, known as the determinant method.
In this paper, we do not use this method but we use the solution matrix to obtain the residue of the Green's functions in the dual field theory on the boundary.

\newcommand{\uIR}{u_{\mathrm{IR}}}
Now, we call the IR boundary of the integral region, $u = u_*$ or $u_0$, as $\uIR$.
In our setup, the physical solution is given by the regular solution at $u=\uIR$.
We consider for the given $\omega$ a set of linearly-independent, regular
solutions $\{\phi_{(J)}(u)\}$ labeled by \(J\).
%In our computation, we set \(\phi^I_{(J)}(u_*)={\delta^I}_J\) without loss of generality.
We can construct a matrix $H$ as follows:
\begin{equation}
	{H(\omega,u)^I}_J = \phi^I_{(J)}(u)\quad
	\text{with}\quad
	{H(\omega,\uIR)^I}_J = {\delta^I}_J.
\end{equation}
The solution matrix is defined by
\begin{equation}
	F_{\omega}(u) = H(\omega, u)H(\omega, u=0)^{-1}.
\end{equation}
This is just a bulk-to-boundary propagator in the Fourier space, i.e., we can write the solution satisfying the IR boundary condition as
\begin{equation}
	\bar{\Phi}_{\omega}(u) = F_{\omega}(u)\varphi_{\omega}.
\end{equation}
Putting this into (\ref{eq:quadratic_action_momentum_positive}), we obtain the onshell quadratic action as
\begin{equation}
	\bar{S}^{(2)}
	= \mathcal{N}
	\int_{0}^{\infty} \frac{\dd{\omega}}{2\pi} \varphi_{-\omega}^{T}\left[
		2 F_{-\omega}^{T} \bar{A}^{H} \partial_{u} F_{\omega}
		+ F^{T}_{-\omega} B^{\dagger}_{\omega} F_{\omega}
	\right]^{u=\uIR}_{u=0} \varphi_{\omega}.
\end{equation}
Note that if $H(u=0)$ is invertible $\lim_{u\to0} F_{\omega}(u) = 1$ by definition.
%The renormalized onshell action is given by $\bar{S}^{(2)} + S_{ct}^{(2)}$.
According to the Lorentzian prescription \cite{Son:2002sd}, we can read the retarded Green's function from the renormalized onshell action as%
\footnote{
	More precisely, the actual Green's function $\mathcal{G}^{R}(\omega)$ in the dual field theory is given by
	$\mathcal{G}^R(\omega)
	%= \mathcal{N} \frac{(2\pi)^2}{\lambda} G^{R}(\omega)
	= \frac{N_{c}}{(2\pi)^2} G^{R}(\omega)$.
	The coefficient comes from the overall factor of the action, $\mathcal{N}$, and the relation between the sources of the observables and the boundary values of the bulk fields given by $(\delta m_q(\omega), \delta A_{\parallel}(\omega))^{T} = \sqrt{\lambda}/(2\pi) \varphi_{\omega}$.
	%The actual Green's function is proportional to $N_{c}$, as same as in \cite{Myers:2007we}, in the end.
}
\begin{equation}
	G^{R}(\omega) =
	- \lim_{\epsilon\to 0} \left[
		\left.\left(
		2 \bar{A}^{H} \partial_{u} F_{\omega}(u) +  \bar{B}_{\omega}^{\dagger}
		\right)\right|_{u=\epsilon} - 2 \bar{C}_{\omega,\epsilon}^{H}
	\right].
	\label{eq:NESS_Green}
\end{equation}
If $H(\omega, u=0)$ is not invertible, $F_{\omega}$ and $G^{R}(\omega)$ are ill-defined.
It implies that $\abs{H(\omega, u=0)}=0$ is a condition for the (quasi)normal mode, where $\abs{H}$ is the determinant of $H$.

According to ref.~\cite{Kaminski:2009dh}, we can obtain the residue of the Green's function, as follows.
The relevant part of the Green's function (\ref{eq:NESS_Green}) can be expressed as
\begin{equation}
	G^{R}(\omega) = - \lim_{u\to 0} 2 \bar{A}^{H} H'(\omega,u)\tilde{H}(\omega,u) \frac{1}{\abs{H(\omega,u)}} + \cdots,
\end{equation}
where we define the adjoint matrix $\tilde{H} = H^{-1} \abs{H}$.
Note that $\tilde{H}$ exists even if $H$ is not invertible.
Close to a single pole of the Green's function denoted by $\omega_{s}$, the determinant of $H$ at $u=0$ as a function of $\omega$ is expanded as 
\begin{equation}
	\abs{H(\omega,0)} = (\omega - \omega_{s}) \left. \frac{\partial}{\partial \omega} \abs{H(\omega,0)} \right|_{\omega=\omega_{s}} + O((\omega-\omega_{s})^2).
\end{equation}
Thus, we obtain the matrix of the residues for $\omega_{s}$, i.e., $G^{R}(\omega) \sim Z/(\omega-\omega_s)$, as
\begin{equation}
	Z = - \lim_{u\to 0} 2 \bar{A}^{H} H'(\omega,u) \tilde{H}(\omega,u) \left. \frac{1}{\partial_{\omega} \abs{H(\omega,u)}} \right|_{\omega_{s}}.
	\label{eq:residue}
\end{equation}
Since $\tilde{H}$ is well-defined and $\partial_{\omega}\abs{H}$ is nonzero at $\omega=\omega_{s}$, this formula gives finite value of the residue.
	
	%%%%%%
	\section{Spectra of mesons}\label{sec:Results}
\begin{figure}[tbp]
		\centering
		\includegraphics[width=14cm]{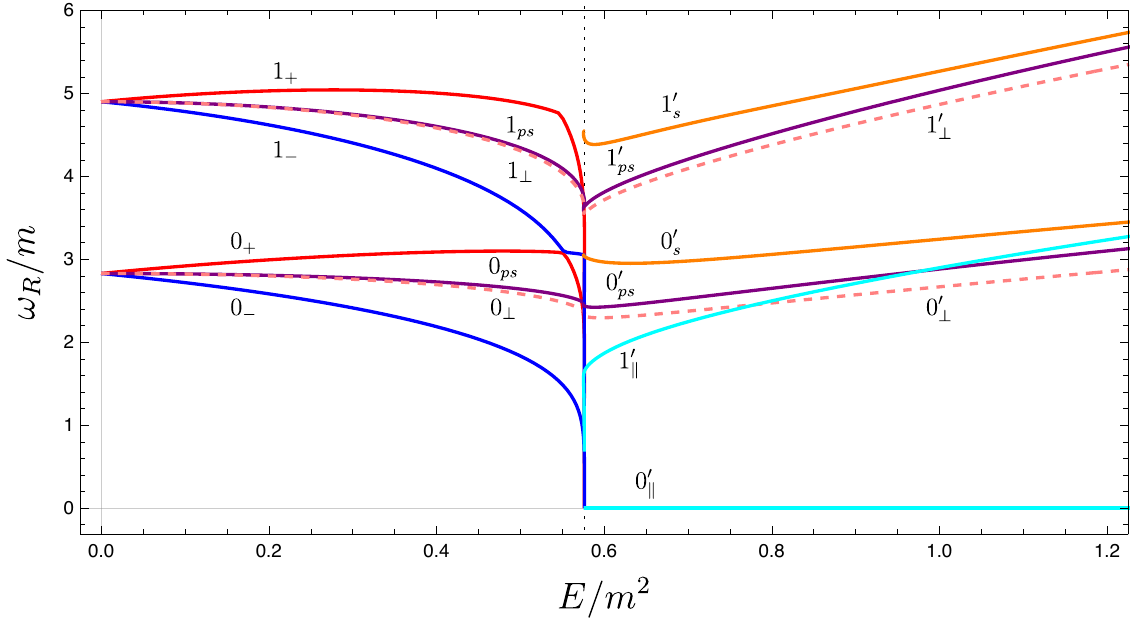}
		\caption{The mass spectra of the lightest two levels ($n=0, 1$) of scalar and vector meson excitations as a function of $E/m^{2}$. The vertical black dotted line denotes the critical value of the electric field, $E_{c}/m^{2}$, and the below/above $E_{c}/m^{2}$ correspond to the meson/dissociation phases. Here, $\chi_{\pm}$ are the coupled modes of the scalar and vector parallel to $E$, denoted by $n_\pm$. Note that they are given by $\chi_{\pm} = w \pm i a_{\parallel}$ at first order in $E$. In large $E$ limit, equivalently massless limit, they are decoupled again and we can identify the scalar $w$ and vector $a_{\parallel}$, respectively. 
	 %That is why we denote the coupled modes as $w (m\to0)$ and $a_{\parallel}(m\to 0)$ in the meson dissociation phase, respectively.
	 That is why we refer to each coupled mode as $n'_s$ and $n'_{\parallel}$ in the dissociation phase.
	 The two perpendicular components of vector $\vec{a}_{\perp}$, denoted by $n_{\perp} (n'_{\perp})$, are always degenerate, and the pseudo-scalar $\psi$ is denoted by $n_{ps} (n'_{ps})$.}
		\label{fig:spectra}
	\end{figure}

	In this section, we study the spectra of the scalar and vector mesons for finite uniform electric fields at zero temperature.
	As studied in \cite{Kruczenski:2003be}, the mass spectra of five modes are degenerate in the vacuum at $E=0$, corresponding to two scalar mesons ($w,\psi$) and three components of the vector meson $a_{i}$ ($i=1,2,3$).
	The electric field couples to each mode and splits the spectra except for the two components of the vector meson perpendicular to the electric field, $\vec{a}_{\perp}$.
	We show the spectra of the scalar and vector mesons for all ranges of the electric field in figures \ref{fig:spectra} and \ref{fig:spectra2}.
	For simplicity, we focus only on the lightest two excitations of the scalar and vector mesons. 
	
	\subsection{Vacuum}
	At zero electric field in the vacuum, the embedding of the probe D$7$-brane is exactly given by %$\theta(u) = \arcsin (mu)$.
$\bar{W}(u) = \sin\theta(u)/u = m$.
	Moreover, every linear perturbation on the D$7$-brane, corresponding to mesonic excitations, can be analytically solved and their mass spectra are shown \cite{Kruczenski:2003be}.
%For convenience, we use Cartesian-like coordinates in a part of the bulk spacetime:
%	\begin{equation}
%		u^{-1} = \sqrt{\rho^{2} + W^2}, \quad \rho= \frac{\cos\theta}{u}, \quad W=\frac{\sin\theta}{u},
%		\label{eq:Cartesian}
%	\end{equation}
%	where the brane embedding is determined by $W=\bar{W}(u) \equiv \sin\theta(u)/u$.
In the vacuum case %$\bar{W}(u) = m$, 
the quadratic action for the perturbations (\ref{eq:effective_action}) reduces to 
	\begin{align}
		S^{(2)} &= - \frac{\mathcal{N}}{2}\int \dd x^{4} \int_0^{1/m} \dd u  \frac{1-m^2u^2}{u^3} g^{ab}\partial_{a}\Phi_{i} \partial_{b} \Phi^{i}  \\
		%&= -\frac{1}{2}\int \dd x^{4} \int_0^\infty \dd \rho \, \rho^{3} \left[ -\frac{1}{(\rho^{2}+m^{2})^{2}} \left(\frac{\partial \Phi_{i}}{\partial t} \right)^{2} + \left(\frac{\partial \Phi_{i}}{\partial \rho} \right)^{2} \right],
	 &= \frac{\mathcal{N}}{2}\int \dd x^{4} \int_0^{1/m} \dd u \, \frac{1-m^2u^2}{u} \left[ \left(\frac{\partial \Phi_{i}}{\partial t} \right)^{2} -(1-m^2u^2) \left(\frac{\partial \Phi_{i}}{\partial u} \right)^{2} \right],
	\end{align}
	where $\Phi = ( w,  a_{\parallel},  \vec{a}_{\perp}, \tilde{\psi})$ with $ \tilde{\psi}\equiv \bar{W}(u) \psi$ and $g_{ab}$ is the induced metric defined by (\ref{eq:induced_metric}).
	The Fourier modes of each field $\Phi_\omega$ obey the same differential equation, 
	\begin{equation}
		{\cal H} \Phi_\omega = \omega^{2}\Phi_\omega,
		 \label{eq:vacuum_deq}
	\end{equation}
	where 
	\begin{equation}
	 {\cal{H}} \equiv -\frac{u}{1-m^{2}u^{2}} \frac{d}{du} \frac{(1-m^{2}u^{2})^{2}}{u} \frac{d}{du}.
	  \label{eq:difop}
	% {\cal H} \equiv - \frac{(\rho^{2}+m^{2})^{2}}{\rho^{3}}
	 %	  \frac{d}{d\rho}\rho^{3}
	 %	  \frac{d}{d\rho} .
	\end{equation}
	To be normalizable on the brane, the eigenfunctions and eigenvalues of ${\cal{H}}$ are given by
	\begin{equation}
		e_{n}(u) = N_{n} m^2u^2 
		 F\left(n+3,-n,2;m^2u^2\right), \quad \omega_{n}^{2}= 4m^{2}(n+1)(n+2),
		 \label{eq:eigen_func}
	\end{equation}
	where $F$ is the Gaussian hypergeometric function and $N_{n}=\sqrt{2(2n+3)(n+1)(n+2)}$ is a normalization factor with $n=0,1,2,\dots$.
	The eigenfunctions are orthonormalized by the inner product 
	\begin{equation}
		\left(f, g \right) \equiv \int^{1/m}_{0} \dd u \frac{1-m^2u^2}{u} f(u) g(u).
		\label{eq:innerproduct}
	\end{equation}

	Thus, 
	the mass spectra of five degrees of freedom for the scalar and vector mesons are degenerate and given by
	\begin{equation}
		\frac{\omega}{m} = 2\sqrt{(n+1)(n+2)} \equiv \frac{\omega_{n}}{m},
		\label{eq:mesonmass}
	\end{equation}
	where $n=0,1,2,\dots$.
	Note that in figure \ref{fig:spectra} the end points of two branches of the mass spectra at $E=0$ correspond to $n=0,1$.
	
  \subsection{Spectra of mesons under the external electric field}
	 Now we look more closely at the spectra of mesons for finite electric field. Before that, we explain the symbols in the spectra for the following discussion. In the meson phase ($E \lesssim E_{c}$), as denoted in figure \ref{fig:spectra}, $n_{\pm}$ correspond to the coupled modes of the scalar meson ($w$) and the vector meson ($a_{\parallel}$) parallel to $E$ at the $n$-th level. The others, $n_{ps}$ and $n_{\perp}$, correspond to the decoupled modes, namely the pseudo-scalar meson ($\psi$) and two perpendicular components of the vector meson ($\vec{a}_{\perp}$), respectively. In the dissociation phase ($E \gtrsim E_{c}$), we explicitly denote the coupled modes of the scalar meson and parallel component of the vector meson as $n'_{s}$ and $n'_{\parallel}$ from the identification at $E \to \infty$ (see later discussion in section \ref{subsec:qnm}). We also denote the decoupled modes as $n_{ps}'$ and $n_{\perp}'$.
  \subsubsection{Normal modes}

	If we switch on the electric field, one scalar meson and one component of vector mesons parallel to the electric field, namely $w$ and $a_{\parallel}$, have been coupled to each other.
	As a result, the electric field lifts the degeneracy of these two modes, the so-called {\it Stark effect}.
	We will analytically study the Stark effect for a weak electric field later.
	%as discussed in \cite{Hashimoto:2014yza}.
	For later convenience, we refer to these two modes as $\chi_{-}$ and $\chi_{+}$ modes\footnote{ As we will show in the section \ref{subsec:stark}, the coupled modes of the scalar and vector component parallel to $E$ can be precisely  diagonalized as $\chi_{\pm} = w\pm i a_{\parallel}$ only at first order in $E$. Although for larger $E$ this combination is no longer valid, we refer to the coupled modes as $\chi_{\pm}$ in the entire region of the meson phase.}, respectively. 
	%As shown in figure \ref{fig:spectra}, the normal frequencies of $\chi_{-}$ modes tend to decrease to zero toward tachyonic instability, whereas those of $\chi_{+}$ apparently approach some finite values as $E\to E_{c}$.
	As shown in figure \ref{fig:spectra}, the normal frequencies of $\chi_{\pm}$ modes (denoted by $n_{\pm}$) are apparently split with increasing the electric field.
	In addition, we find that the spectra of the other mesons, $\psi$ and $\vec{a}_{\perp}$, also are slightly split by the electric field.
	We will also analytically discuss this split in the later section.
	
	Interestingly, we find that the $\chi_{+}$ mode with $n=0$ (the $0_+$ mode) and the $\chi_{-}$ mode with $n=1$ (the $1_-$ mode) nearly collide at $E/m^{2}\approx 0.55$ but bounce back without crossing.
	This behavior is considered as the {\it avoided crossing} or {\it level repulsion}.
	The level repulsion implies that the mixing of scalar and vector mesons is essential.
	If these two modes could be decoupled by changing and redefining the variables, no interactions of those eigenvalues in the mass spectra are expected.
	We will discuss the avoided crossing in depth in the later section.
	
\begin{figure}[tbp]
			\centering
			\includegraphics[width=12cm]{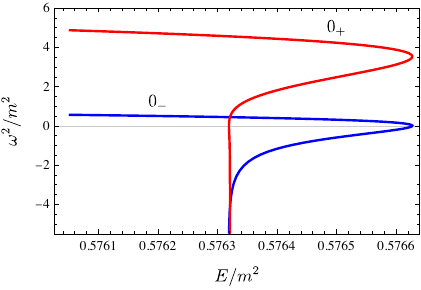}
			\caption{$\omega^{2}/m^{2}$ as a function of $E/m^{2}$ for the $0_{-}$ and $0_{+}$ modes close to the critical electric field.}
			\label{fig:unstable}
	\end{figure}

	In order to focus on actual behaviors close to the critical electric field, we show $\omega^{2}/m^{2}$ for the $0_{-}$ and $0_{+}$ modes in figure \ref{fig:unstable}.
	For each mode the multivaluedness of the mass spectrum at a given $E/m^{2}$ stems from the non-uniqueness and discrete self-similarity of the embedding solutions in the vicinity of the critical embedding (see figure~\ref{fig:EJplot}) as discussed in section \ref{sec:critical}.
	The appearance of possible solutions for the same parameter, namely, branching of the solutions implies that linear stability may change.
	Indeed, it turns out that the $0_-$ mode becomes tachyonic at 
	$E/m^2 \approx 0.57663$, where the mass spectrum, as well as the background parameter, becomes two-valued.
	Such a tachyonic instability in the meson phase was also observed with finite temperature in \cite{Mateos:2007vn}, instead of the electric field.
	In summary, the turnovers in the $E$-$J$ and $E$-$c$ curves (figure \ref{fig:EJplot}), which originate from the discrete self-similarity of the brane embedding around the critical embedding solution, correspond to the onset of the tachyonic instability as well as the appearance of the static perturbation (figure \ref{fig:unstable}).
	It is notable that the subsequent tachyonic mode is the $0_+$ mode but not the $1_-$ mode, whose mass tends to decrease to zero for a small $E$.
	The avoided crossing has resulted in this, as will be shown later in figure~\ref{fig:avoided_crossing}.

	\subsubsection{Quasinormal modes} \label{subsec:qnm}
	
	Above around the critical value $E_{c}/m^{2}\approx 0.57632$, the mesons are dissociated by the electric field and %the vector meson condensate
the electric current density becomes finite in the background.
	Due to the dissociation, the fluctuations show no longer normal modes but quasi-normal modes whose frequencies become complex numbers with an imaginary part.
	In \cite{Ishigaki:2021vyv}, the authors have mainly focused on the purely imaginary mode to elucidate the dynamical stability of the non-equilibrium steady state.
	For our present purpose, we focus on the real part of the lower (quasi-)normal modes to study the mass spectra of the scalar and vector mesons.

\begin{figure}[tbp]
	 \centering
	 \includegraphics[width=14cm]{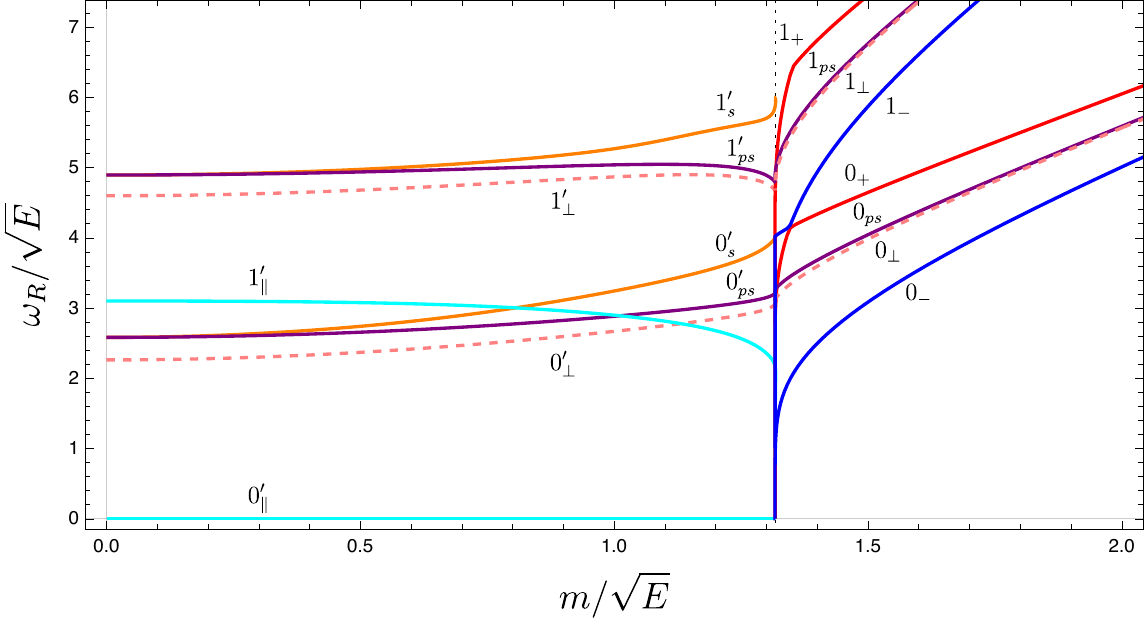}
	 \caption{The mass spectra of the lightest two scalar and vector meson excitations as a function of $m/\sqrt{E}$. 
	 %The only difference with figure \ref{fig:spectra} is that we use $m/\sqrt{E}$ as the horizontal axis instead of $E/m^{2}$
	 This figure is essentially identical to figure \ref{fig:spectra} but the only difference is that we use $\omega_R/\sqrt{E}$ as the vertical axis instead of $\omega_R/m$ in order to be appropriately scaled in the massless limit.
	 The vertical dotted line denotes the critical value, above(below) which the system is in the insulating(conducting) phase.
	 }
	 \label{fig:spectra2}
	\end{figure}

	For $E\gtrsim E_{c}$, we find that two kinds of the decoupled modes, $\psi$ and $\vec{a}_{\perp}$, are continuously connected at $E=E_{c}$.
	Besides, we find that the other two coupled modes behave differently: One mode, denoted by $n'_s$, approaches a normal mode as $E\to E_{c}$, whose frequency has no imaginary part; the other mode, denoted by $n'_{\parallel}$, becomes a purely imaginary mode without the real part and then becomes unstable.
	It should be noted that we cannot distinguish between the $w$ and $a_{\parallel}$ modes at finite $E$.
	However, those quasi-normal frequencies can be obtained independently because the fluctuations $w$ and $a_{\parallel}$ are decoupled again in the massless limit.
	(We show some lower quasi-normal frequencies for $a_{\parallel}$, $\vec{a}_{\perp}$, and $w$ in the massless limit in Table \ref{tab:massless_QNMs}.)
	That is why we have referred to each of the coupled modes that approaches $w$ and $a_\parallel$ in the massless limit as $n'_{s}$ and $n'_{\parallel}$, respectively.
	Indeed, we find that in the limit of $m\to0$ the quasi-normal frequency of the $n'_s$ mode with $n=0$ approaches $\omega/\sqrt{E} \approx 2.58 -1.44 i$ and that of $n'_\parallel$ with $n=0$ does $\omega / \sqrt{E} \approx - \sqrt{6} i$.
	To see this, we show the mass spectra as a function of $m/\sqrt{E}$ in figure \ref{fig:spectra2}.
	Note that in this figure the region left (right) $m/\sqrt{E_{c}}$ corresponds to the dissociate (meson) phase and the massless limit corresponds to the left endpoints.

	\begin{figure}[tbp]
	 \centering
	 \includegraphics[width=14cm]{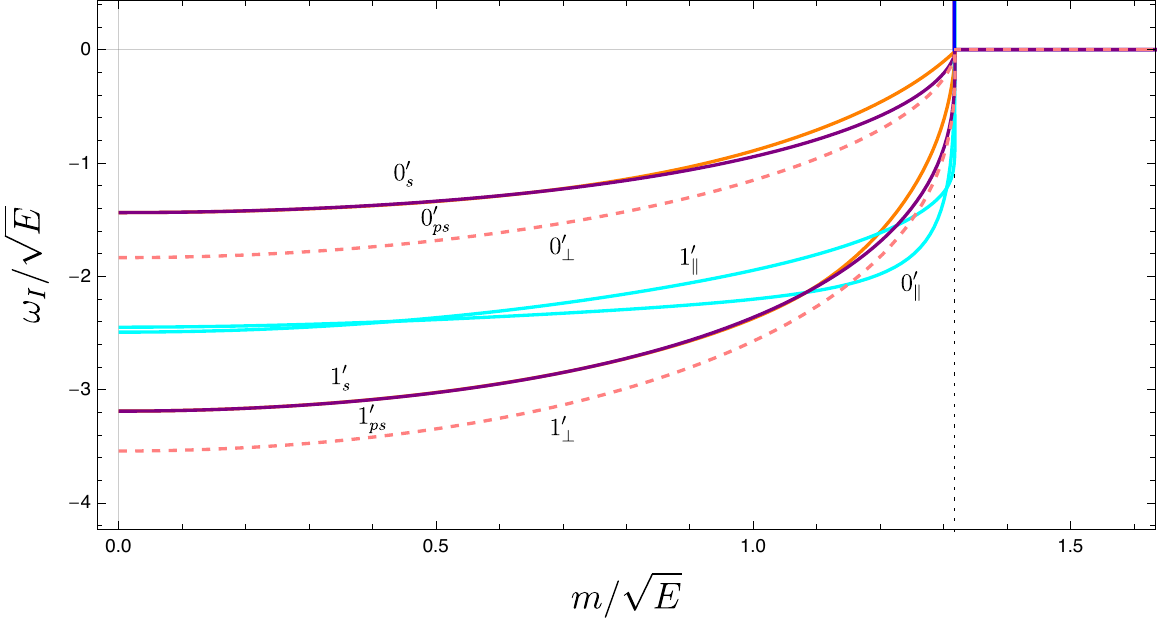}
	 \caption{The imaginary part of the mass spectra of the lightest two scalar and vector meson excitations as a function of $m/\sqrt{E}$. 
	 }
	 \label{fig:spectra2_im}
	\end{figure}
	\begin{table}[tbp]
		\centering
		%\label{tab:massless_QNMs}
		\caption{
			The quasi-normal frequencies $\omega_n/\sqrt{E}$ for $a_{\parallel}$, $\vec{a}_{\perp}$, and $w$ and $\psi$ in the massless limit.
			%(pseudospecral $N=40$).
		}
	 \label{tab:massless_QNMs}
		\begin{tabular}{c | c c c}
			& $n'_{\parallel}$ & $n'_{\perp}$ & $n'_s , n'_{ps}$\\\hline
			$0$&
			$-2.44949 i (= - \sqrt{6} i)$ &
			$\pm 2.26082\, -1.83585 i$ & $\pm 2.58134-1.43819 i$\\
			$1$&
			$\pm 3.0999-2.49329 i$ & $\pm 4.59708\, -3.54243 i$ & $\pm 4.89226-3.19041 i$\\
			$2$&
			$\pm 5.63587\, -4.31466 i$ & $\pm 6.93715\, -5.25984 i$ & $\pm 7.22005-4.91746 i$\\
			$3$&
			$\pm 8.03395\, -6.06627 i$ & $\pm 9.27658-6.97908 i$& $\pm 9.55221-6.64093 i$\\
			$4$&
			$\pm 10.3988-7.79697 i$ & $\pm 11.6164-8.70476 i$& $\pm 11.8997-8.36464 i$
		\end{tabular}
	\end{table}

\begin{figure}[tbp]
		\centering
		\includegraphics[width=12cm]{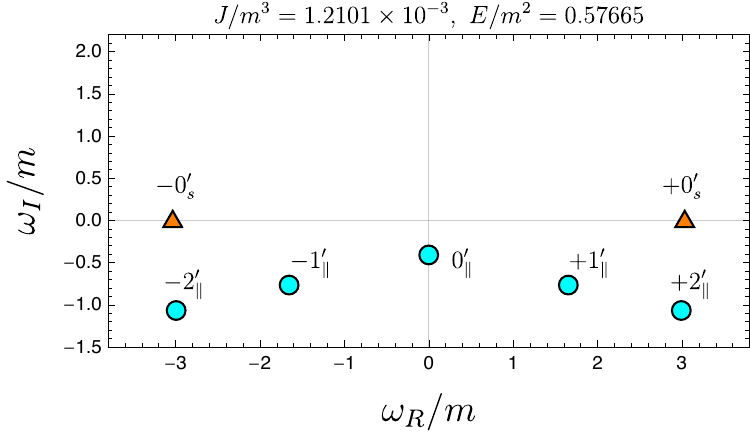}
		\caption{The frequencies of the QNMs on the complex plane with the identifications of mesons in the vicinity of the critical electric field. The figure corresponds to figure 7 in \cite{Ishigaki:2021vyv} with the modifications.}
		\label{fig:qnm}
	\end{figure}

	Before closing this subsection let us mention the relation of the meson spectrum in the dissociation phase to the previous analysis for studying the dynamical instability \cite{Ishigaki:2021vyv}. Although we mainly focused on the purely imaginary modes in \cite{Ishigaki:2021vyv}, we already showed other meson excitations in the QNMs analysis. We show the quasi-normal frequencies on the complex plane with the identifications of mesons in figure \ref{fig:qnm}, corresponding to figure 7 in \cite{Ishigaki:2021vyv}. 
	As shown, the $0'_{s}$ modes appear near the real axis in the vicinity of the critical electric field ($J\to0$ in the dissociation phase), and continuously connect to the normal modes in the meson phase beyond the critical embedding as can be seen in figure \ref{fig:spectra} or \ref{fig:spectra2}.

	%%%%%%
	\subsection{Stark effect of coupled modes for small $E$} \label{subsec:stark}
	As discussed above, at $E=0$ all the fluctuations on the D$7$-brane are decoupled and both the mode functions and the normal frequencies corresponding to the degenerate mass spectra can be analytically represented.
	%As discussed above, at $E=0$ all the fluctuations are decoupled and the normal frequencies correspond to the degenerate mass spectra given by (\ref{eq:mesonmass}) with $n=0$.	
	In general, for finite $E$, the fluctuations of $w$ and $a_{\parallel}$ are coupled via the electric field, and the mode functions can be no longer analytical.
	In this subsection, we analytically study the Stark effect for small electric fields in a perturbative approach as discussed in Appendix of ref.~\cite{Hashimoto:2014yza}.
	
	We write the static solutions for the linear order in the electric field as
	\begin{equation}
		\bar{W}(u) = \frac{\sin\theta(u)}{u} = m+{\cal{O}}(E^{2}), \quad \bar{A}_{x} = -Et.
		\label{eq:staticsol}
	\end{equation}
	We consider the fluctuations of the static solutions: $W(t,u)=\bar{W}(u)+w(t,u)$ and $A_{x}(t,u) = \bar{A}_{x} + a_{\parallel}(t,u)$.
	The quadratic action for the fluctuations up to the first order in $E$ is given by
	\begin{equation}
	 \begin{aligned}
	  S^{(2)}_{w, a_{\parallel}}%&=-\frac{1}{2}\int\dd t \int^{1/m}_{0} \dd u \frac{1-m^{2}u^{2}}{u} \left[\cdots\right]\\
	  &= \frac{\mathcal{N}}{2}\int\dd t \int^{1/m}_{0} \dd u \frac{1-m^{2}u^{2}}{u} \left[(\dot{w}^2 + \dot{a}_{\parallel}^2)
	  - (1-m^2u^2)(w'^2 + {a'_{\parallel}}^2)
	  + 2Emu^2 (w \dot{a}_{\parallel} - \dot{w}a_{\parallel})\right]\\
	  &= \frac{\mathcal{N}}{2}\int\dd t \int^{1/m}_{0} \dd u \frac{1-m^{2}u^{2}}{u}\left[ \dot{\chi}_{+}\dot{\chi}_{-} -(1-m^{2}u^{2})\chi_{+}' \chi_{-}' -2 i E m u^{2} \left( \dot{\chi}_{+}\chi_{-} -\chi_{+}\dot{\chi}_{-} \right)\right],	  
	 \end{aligned}
	\end{equation}
	where $\chi_{\pm}= w\pm i a_{\parallel}$.
	Decomposing the fields into the Fourier modes as $\chi_{\pm}=\int \frac{\dd\omega}{2\pi} \chi^{\pm}_{\omega}(u) e^{-i\omega t}$, we obtain the decoupled equations 
	\begin{equation}
		\left( {\cal{H}} \pm 4 E \omega m u^{2} \right) \chi_{\omega}^{\pm} = \omega^{2} \chi_{\omega}^{\pm},
	\end{equation}
	where the operator ${\cal{H}}$ is defined by~(\ref{eq:difop}).
	Thus, it turns out that at linear order in $E$ two modes $w$ and $a_{\parallel}$ should be mixed, while the mixed modes $\chi_\pm$ are decoupled.
%	\begin{equation}
%		{\cal{H}} \equiv -\frac{u}{1-m^{2}u^{2}} \frac{d}{du} \frac{(1-m^{2}u^{2})^{2}}{u} \frac{d}{du}.
%		\label{eq:difop}
%	\end{equation}
%	The eigenfunctions and eigenvalues of ${\cal{H}}$ are given by
%	\begin{equation}
%		e_{n} = N_{n} m^{2}u^{2} F(n+3,-n,2;m^{2}u^{2}), \quad \omega_{n}^{2}= 4m^{2}(n+1)(n+2),
%	\end{equation}
%	where $F$ is the Gaussian hypergeometric function and $N_{n}=\sqrt{2(2n+3)(n+1)(n+2)}$ is a normalization factor with $n=0,1,2,\cdots$.
%	Defining the inner product as
%	\begin{equation}
%		\left(f, g \right) \equiv \int^{1/m}_{0} \dd u \frac{1-m^{2}u^2}{u} f(u) g(u),
%	\end{equation}
%	the eigenfunction is orthonormalized.
A complete set of the eigenfunctions of the operator $\cal{H}$ is given by (\ref{eq:eigen_func}).
	%Note that $(e_{n}, m^{2}u^{2}e_{n}) = 1/2$ is satisfied for any $n$.
	If we consider the shift of the eigenvalues, $\delta \omega_{n}$, up to the linear order of $E$, it is given by the
	\begin{equation}
		\delta\omega_{n} = \left(e_{n}, {\Delta\cal{H}} e_{n} \right)= \frac{2E}{m}\left( e_{n}, m^{2}u^{2} e_{n}\right) = \frac{E}{m},
	\end{equation}
	where $(\cdot,\cdot)$ is the inner product defined in (\ref{eq:innerproduct}). Here, ${\Delta\cal{H}}=2E mu^{2}$ is considered as the perturbative contribution to the eigenvalues and given from the eigenequation in the linear order of $E$, that is $\left({\Delta\cal{H}} \pm \delta\omega_{n} \right) \chi_{\omega}^{\pm}=0$.
	Note that $(e_{n}, m^{2}u^{2}e_{n}) = 1/2$ is satisfied for any $n$.
	%Thus, the meson spectra for $E=0$ corresponds to $\omega_{n}^{\pm} = 2\sqrt{(n+1)(n+2)}$ \cite{Kruczenski:2003be}.
	Thus, the shifts of the spectra for small $E$ are given by $\delta\omega_{n}^{\pm} = \pm E/m$, which are numerically reproduced in figure \ref{fig:spectra}.
	 This means that the first-order Stark effect occurs in the coupled modes. 
	
	%%%%%%
	\subsection{Split in decoupled modes for small $E$}
\begin{figure}[tbp]
		\centering
		\includegraphics[width=10cm]{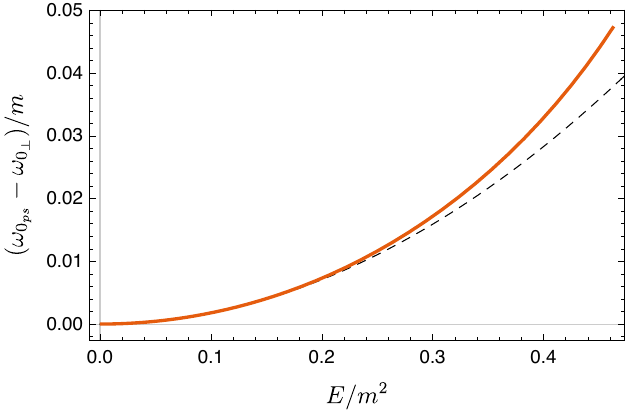}
		\caption{The relative frequency shift between $ \vec{a}_{\perp}$ and $ \psi$ as a function of $E/m^{2}$. Here, $\omega_{0_{\perp}}$ and $\omega_{0_{\rm ps}}$ denote the normal frequencies of $ \vec{a}_{\perp}$ and $\psi$ with $n=0$, respectively. The solid and dashed lines denote the numerical results and analytical prediction up to the order of $E^{2}$.}
		\label{fig:spectra_difference}
	\end{figure}

	Here, we study the mass split of the decoupled modes, $\psi$ and $\vec{a}_{\perp}$, for small $E$.
	%For convenience, we use the Cartesian coordinate
%	\begin{equation}
%		u^{-1} = \sqrt{\rho^{2} + \bar{W}(\rho)^{2}}, \quad \rho= \frac{\cos\theta}{u}, \quad \bar{W}(\rho)=\frac{\sin\theta}{u},
%		\label{eq:Cartesian}
%	\end{equation}
%	where $\bar{W}(\rho)$ determines the brane embedding as in the above discussion.
%	At $E=0$, the quadratic action for the perturbations are written as
%	\begin{align}
%		S^{(2)} &= - \frac{1}{2}\int \dd x^{4} \int \dd \rho  \frac{\rho^{3}}{\rho^{2}+m^{2}} g^{ab}\partial_{a}\Phi_{i} \partial_{b} \Phi^{i}  \\
%		&= -\frac{1}{2}\int \dd x^{4} \int \dd \rho \, \rho^{3} \left[ -\frac{1}{(\rho^{2}+m^{2})^{2}} \left(\frac{\partial \Phi_{i}}{\partial t} \right)^{2} + \left(\frac{\partial \Phi_{i}}{\partial \rho} \right)^{2} \right],
%	\end{align}
%	where $\Phi = ( w,  a_{\parallel},  \vec{a}_{\perp}, \tilde{\psi})$ with $ \tilde{\psi}\equiv \bar{W}(\rho) \psi$.
%	For convenience, we introduce new coordinates, $z=m / \sqrt{\rho^{2}+m^{2}}$ and $\tilde{t}=mt$.
%	Then, the action can be written as
%	\begin{align}
%		S^{(2)} = \frac{m}{2}\int \dd \tilde{t} \int_{0}^{1} \dd z \frac{1-z}{z^{2}}\left[ \left( \frac{\partial\Phi_{i}}{\partial \tilde{t}} \right)^{2} - (1-z^{2}) \left( \frac{\partial\Phi_{i}}{\partial \rho} \right)^{2} \right].
%	\end{align}
	At $E=0$, the quadratic action for the perturbations of $\tilde{\psi} \equiv m\psi$ and $\vec{a}_{\perp}$ are the same.
	For finite $E$, we find that the forms of the kinetic term for $ \vec{a}_{\perp}$ and $\tilde{\psi}$ are identical, but a mass term for $\tilde{\psi}$ is added on the order of $E^{2}$ as 
%	\begin{align}
%		\delta S^{(2)}_{\tilde{\psi}} = -E^{2} \int \dd t \dd \rho \frac{\rho^{3}}{(\rho^{2} +m^{2})^{3}} \tilde{\psi}^{2}  =- \frac{E^{2}}{m^{3}} \int \dd \tilde{t} \int_{0}^{1} \dd z \, z(1-z) \tilde{\psi}^{2}.
%	\end{align}
	\begin{align}
		\delta S^{(2)}_{\tilde{\psi}} = - \mathcal{N} E^{2} \int \dd t \int_0^{1/m} \dd u \, u(1-m^2u^2) \tilde{\psi}^{2} .
	\end{align}
	Following the discussion in the Stark effect, we find that the relative mass shift is given by
	\begin{equation}
		\delta m_{n}^{2}= \frac{2E^{2}}{m^{2}} \left(e_{n}, m^2u^{2}e_{n} \right) = \frac{E^{2}}{m^{2}}.
	\end{equation}
	Since the relative mass shift corresponds to the relative frequency shift as $\delta m_{n}^{2} = (\omega_{n}+ \delta\omega_{n})^{2} -\omega_{n}^{2} \approx  2\omega_{n}\delta \omega_{n} $, we find
	\begin{equation}
		\frac{\delta \omega_{n}}{m} = \frac{E^{2}}{2 \omega_{n} m^{3}} = \frac{1}{4\sqrt{(n+1)(n+2)}}\frac{E^{2}}{m^{4}}. 
		\label{eq:shift}
	\end{equation}
	In figure \ref{fig:spectra_difference}, we show the relative frequency shift between $\vec{a}_{\perp}$ and $\psi$ as a function of $E/m^{2}$.  
	The solid and dashed lines denote the numerical results and analytical prediction in (\ref{eq:shift}) up to the order of $E^{2}$, that is $E^{2}/ (4\sqrt{2} m^{4})$.
	As shown, we find that the numerical results of the relative frequency are well matched to the analytical prediction for small $E/m^{2}$.

	Note that since the background brane embedding is also affected in the order of $E^{2}$ as in (\ref{eq:staticsol}), we discussed only the relative frequency shift between $ \vec{a}_{\perp}$ and $\psi$.
	If one wants to study the absolute shifts of these modes that are the second-order effects in $E$, we have to consider the contributions of %the background solutions, i.e. $\bar{W} = m + {\cal{O}}(E^{2})$.
	a perturbed background solution in the order of $E^2$, i.e. 
	$\bar{W}(u) = m + E^2 \delta W(u) + {\cal{O}}(E^{4})$.
	The absolute shift of the decoupled mode $\psi$ was discussed in \cite{Erdmenger:2007bn}.

	%%%%%%%
	\subsection{Avoided crossing}
\begin{figure}[tbp]
		\centering
		\includegraphics[width=7cm]{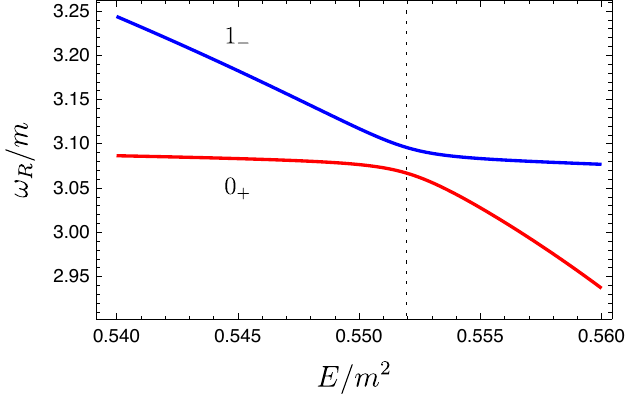}
		\includegraphics[width=7cm]{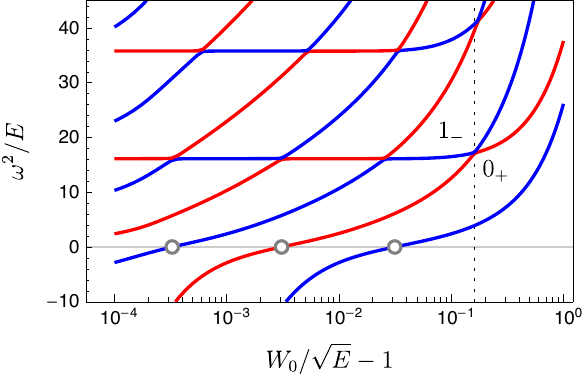}
		\caption{
		The left panel shows the avoided crossing between the $0_{+}$ and $1_{-}$ modes in the spectrum.
		The vertical line shows the nearest point of the two levels: $\Enear/m^2 = 0.55196$.
		The right panel shows that, in the vicinity of the critical electric field, the avoided crossing repeatedly happens toward the critical embedding. For this purpose, we plot $\omega^{2}/E$ as a function of $W_{0}/\sqrt{E}-1$ on a semi-log scale.
		The vertical line shows the corresponding value of $W_{0}/\sqrt{E}-1$ to $\Enear/m^2$.
		%The small circles denote the locations where the static perturbation realizes, i.e., the modes with $\omega^2=0$.
	The small circles on $\omega^2=0$ denote the onsets of the tachyonic instability, at which static perturbations appear.
		}
		\label{fig:avoided_crossing}
	\end{figure}

	%As mentioned above, we find the avoided crossing between the $\chi_{+}$ mode with $n=0$ (the $0_{+}$ mode) and the $\chi_{-}$ mode with $n=1$ (the $1_{-}$ mode) at $E/m^{2}\approx 0.55 $ as shown in figure \ref{fig:spectra}. 
	As mentioned above, we find the avoided crossing between the $0_{+}$ and $1_{-}$ modes at $E/m^{2}\approx 0.55 $ in figure \ref{fig:spectra}.
	Zooming in on that region, these two modes are obviously repulsive in the spectra shown in the left panel of figure \ref{fig:avoided_crossing}.
	We define the point at which the two levels are nearest as $\Enear/m^2 = 0.55196$.
	This point can be said as a center of the avoided crossing between the $0_{+}$ and $1_{-}$ modes.
	In the vicinity of the critical electric field, we also find that the avoided crossing repeatedly happens between different levels. To see this, we plot $\omega^{2}/E$ as a function of $W_{0}/\sqrt{E}-1$ on a semi-log scale in the right panel of figure \ref{fig:avoided_crossing}. Note that we have used $W_{0}/\sqrt{E}-1$ as a parameter to uniquely describe the background without being multivalued. The embedding approaches the critical embedding with $W_0 \to \sqrt{E} + 0$, corresponding to the critical electric field. The plot also implies that the lower modes become tachyonic in order as a result of the avoided crossing.%
\footnote{
This, together with the result in figure \ref{fig:scale}, indicates that 
the $n_-$ mode will become tachyonic at the $(2n+1)$-th turning point, and 
the $n_+$ mode will at the $(2n+2)$-th turning point.
}
	
	To analyze the avoided crossing in-depth, we compute residues of the retarded Green's function by using (\ref{eq:residue}).
	We find that the residue matrix has characteristic behavior in the center of avoided crossing point, as we shall show later.
	
	For later discussion, we consider the residues and Green's
	function in the vacuum at $E=0$.
	As mentioned, in the vacuum the five degrees of freedom for the
	mesons (two scalar and three components of vector) are
	degenerate and are governed by (\ref{eq:vacuum_deq}).
	In the bulk, the homogeneous solution for each field is 
	\begin{equation}
	 \Phi_\omega(u) = \frac{{}_2F_1(\alpha_+ , \alpha_- , 2; 1-m^2u^2)}
	  {{}_2F_1(\alpha_+ , \alpha_- , 2; 1)} ,
	  \quad \alpha_\pm = 
	  \frac{1}{2} \pm \frac{1}{2}\sqrt{1+ \frac{\omega^2}{m^2}} ,
	\end{equation}
	where $\Phi_\omega(u)$ is required to be regular at $u=1/m$ and $\Phi_\omega(u)=1$ at $u=0$.
	Thus, the retarded Green's function in the boundary 
	(see in appendix B of~\cite{Myers:2007we}, for example) is 
	\begin{equation}
	 %\begin{aligned}
	  G^{R}(\omega)
	  =  \frac{\omega^2}{4} \left(-1 + H_{\alpha_+} + H_{\alpha_-}\right)
	  % + \frac{\omega^2}{4}\log m^2\epsilon^2 
	  = m^2\omega^2 \sum_{n=0}^\infty
	  \frac{(2n+3)\omega^2}{\omega^2_n(\omega^2 - \omega^2_n)},
	 %\end{aligned}	 
	\end{equation}
	where $H_s$ is the harmonic number.
	%The poles are given by $\alpha_- = -n$ for $n=0,1,2,\ldots,$
	The poles are located at $\omega = \pm \omega_{n}$ given by (\ref{eq:mesonmass}), and the corresponding residues are
	\begin{equation}
		Z_{n} = \pm  m^{3}(2n+3)\sqrt{(n+1)(n+2)}.
		\label{eq:residueE0}
	\end{equation}

\begin{figure}[tbp]
		\centering
		\includegraphics[height=4.55cm]{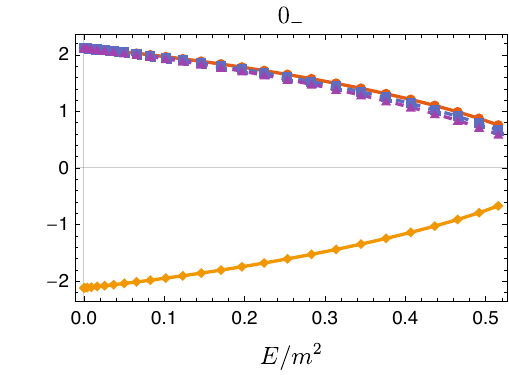}
		\includegraphics[height=4.55cm]{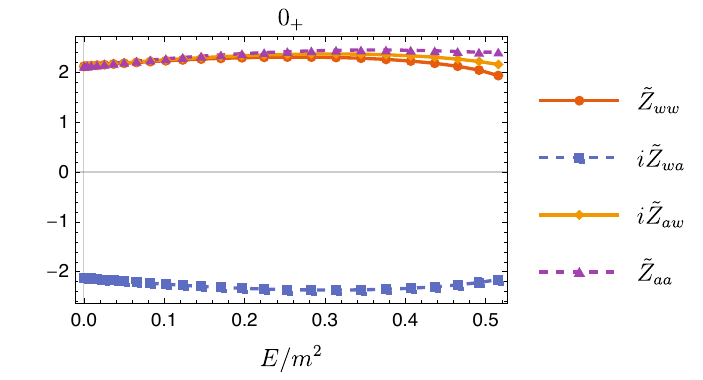}
		\caption{The residues as a function of $E/m^{2}$ for the $\chi_{-}$ (left panel) and $\chi_{+}$ (right panel) modes with $n=0$.
		$\tilde{Z}_{IJ} = Z_{IJ}/m^3$ corresponds to each component of the matrix residue: $\tilde{Z}_{ww}$ and $\tilde{Z}_{aa}$ denote the diagonal part and $\tilde{Z}_{wa}$ and $\tilde{Z}_{aw}$ denote the off-diagonal part, respectively. At $E/m^{2}=0$, the absolute values of the components reach $3 \sqrt{2}/2 \approx 2.1213$.}
		\label{fig:residue1}
	\end{figure}

	Let us see the behaviors of the residues for finite $E$.
	Figure \ref{fig:residue1} shows the residues as a function of $E/m^{2}$ for the $0_{-}$ mode (the left panel) and the $0_{+}$ mode (the right panel).
	Note that we show the four components of the residue matrix because $w$ and $a_{\parallel}$ are coupled, and the Green's function is given by a $2\times2$ matrix.
	Here, $Z_{ww}$ and $Z_{aa}$ denote the diagonal part and $Z_{wa}$ and $Z_{aw}$ denote the off-diagonal part, respectively.
	The numerical results imply that the linear response has the following contributions in the Fourier space for small $E$:
	\begin{equation}
		\begin{pmatrix}
			w^{(+)}(\omega) \\ a^{(+)}(\omega)
		\end{pmatrix}
		\approx
		\left\{
		\frac{(3\sqrt{2}/2) m^3}{\omega - (2\sqrt{2} m - \frac{E}{m})}
		\begin{pmatrix}
			1 & - i\\
			i & 1
		\end{pmatrix}
		+ \frac{(3\sqrt{2}/2) m^3}{\omega + (2\sqrt{2} m + \frac{E}{m})}
		\begin{pmatrix}
			1 & i\\
			- i & 1
		\end{pmatrix}
		\right\}
		\begin{pmatrix}
			w^{(0)}(\omega) \\ a^{(0)}(\omega)
		\end{pmatrix}
		+ \cdots,
	\end{equation}
	where $w^{(+)}, a^{(+)}$ and $w^{(0)}, a^{(0)}$ are responses and sources for each perturbation, respectively.
	Note that the solutions are expanded into $w(\omega, u) = w^{(0)}(w) + \frac{w^{(+)}}{2}u^2 + \order{u^4}$ and $a_{\parallel}(\omega, u) = a^{(0)}(\omega) + \frac{a^{(+)}}{2} u^2 + \order{u^4}$.
	The ellipsis denotes the contributions from the other poles and the regular part.
	In the above equation, the first and second terms represent contributions from the $0_{-}$ and $0_{+}$ modes, respectively.
	If we take a limit of $E\to0$, the two modes degenerate.
	Then, we obtain
	\begin{equation}
		\begin{pmatrix}
			w^{(+)}(\omega) \\ a^{(+)}(\omega)
		\end{pmatrix}
		=
		\frac{3\sqrt{2}m^3}{\omega - 2\sqrt{2} m}
		\begin{pmatrix}
			1 & 0\\
			0 & 1
		\end{pmatrix}
		\begin{pmatrix}
			w^{(0)}(\omega) \\ a^{(0)}(\omega)
		\end{pmatrix}
		+ \cdots.
		\label{eq:residues_decoupling}
	\end{equation}
	The residue becomes diagonal because the perturbations are decoupled at $E=0$.
	This is expected behavior, and the value of residue agree with the analytic result (\ref{eq:residueE0}) at $E=0$.
	
\begin{figure}[tbp]
		\centering
		\includegraphics[height=4.55cm]{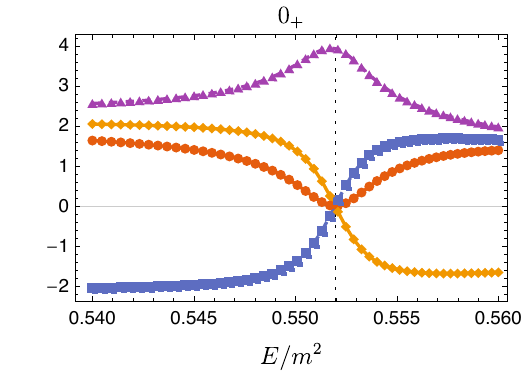}
		\includegraphics[height=4.55cm]{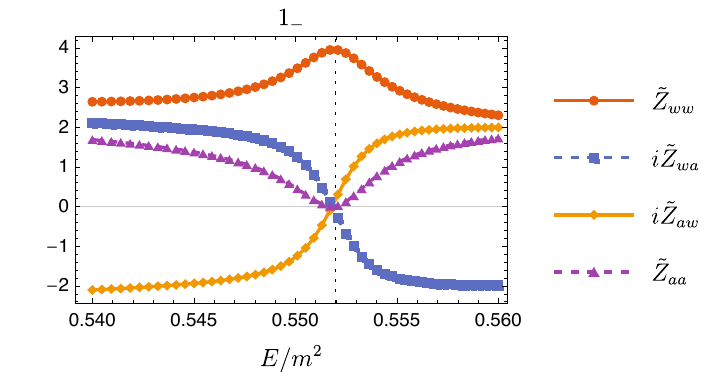}
		\caption{The residues as functions of $E/m^{2}$ for the $0_{+}$ mode (left panel) and the $1_{-}$ mode (right panel).
		The vertical dotted lines show $\Eac/m^2 = 0.55195$ corresponding to the center of the avoided crossing.
		%It was obtained from where $Z_{wa}$ across zero for $0_{+}$.
		}
		\label{fig:residue2}	
	\end{figure}

	Now we show the behaviors of the residues near the avoided crossing in figure \ref{fig:residue2}.
	Interestingly, we find the point at which only one component of the residue matrix remains finite, and the others vanish in both the $0_{+}$ and $1_{-}$ modes.
	We obtain this point as $\Eac/m^2 = 0.55195$.
	%by finding where $Z_{wa}$ of $0_{+}$ mode goes across zero.
	At $E=\Eac$, the numerical results imply
	\begin{equation}
		\begin{pmatrix}
			w^{(+)}(\omega) \\ a^{(+)}(\omega)
		\end{pmatrix}
		=
		\left\{
			\frac{Z^{(0_{+})}_{aa}}{\omega - \omega_{0_{+}}}
			\begin{pmatrix}
				0 & 0\\
				0 & 1
			\end{pmatrix}
			+
			\frac{Z^{(1_{-})}_{ww}}{\omega - \omega_{1_{-}}}
			\begin{pmatrix}
				1 & 0\\
				0 & 0
			\end{pmatrix}
		\right\}
		\begin{pmatrix}
			w^{(0)}(\omega) \\ a^{(0)}(\omega)
		\end{pmatrix}
		+ \cdots,
		\label{eq:residues_at_avoided_crossing}
	\end{equation}
	where $\omega_{1_-}$ and $\omega_{0_{+}}$ denote the frequencies at $E=\Eac$.
	$Z^{(1_{-})}_{ww}$ and $Z^{(0_{+})}_{aa}$ denote the only remaining component of each matrix of the residues.
	$\omega_{1_{-}}$ and $\omega_{0_{+}}$ are very close, but slightly different values due to the avoided crossing
	(see figure \ref{fig:avoided_crossing}).
	As shown in figure \ref{fig:avoided_crossing}, $\Eac$ is almost the same as $\Enear$, which is a center of the avoided crossing between the $0_{+}$ and $1_{-}$ modes.
	%Our results of $\Enear$ and $\Eac$ are same up to four digits.
	Our results indicate good agreement between $\Enear/m^2$ and $\Eac/m^2$ up to four digits.
	%Actually, it agrees with the nearest point of the spectra between $0_{+}$ and $1_{-}$.
	From the above observations, we expect that the vanishing point of most components of the residue matrix corresponds to the nearest point for each avoided-crossing point.

	\section{Conclusion and discussion} \label{sec:Conclusion}
	In this paper, by using the D3/D7 model we have elucidated the mass spectrum of mesons over the whole range of external electric fields at zero temperature;
 	the zero electric field corresponds to the supersymmetric vacuum state, whereas the large electric field limit to the massless quark limit.	
	%In this paper, we analyze the mass spectrum of mesons at a finite electric field by using the D3/D7 model.
	In the dual field theory the system contains charged particles in the fundamental representation strongly interacting with the heat bath which is consisted of fields in the adjoint representation. %the gauge particles.
	For small electric fields the mesons exist stably as the quark-antiquark bound states, while for large electric fields they are dissociated into each particle that behaves as a charge carrier.
	In terms of excitations of the D$7$-brane, the former are described by normal modes, the latter by quasinormal modes. 
	%The charged particles exist as the meson states for small electric field, whereas they are dissociated into each particle for large electric field.
	At zero electric field, it is known that the mass spectra are degenerate in five degrees of freedom: two scalar mesons and three components of one vector meson, corresponding to fluctuations on the D$7$-brane for the two embedding functions and the worldvolume gauge field.
	%At zero electric field, the five mesons are degenerate in the mass spectra, corresponding to the normal modes for the fluctuations on the D7-brane: two scalars and three components of gauge fields \cite{Kruczenski:2003be}.

	When an external electric field is applied, one scalar meson and one component of the vector meson along the electric field are coupled, while the other mesons remain decoupled.
	As a result, we observe that the mass spectra of the two mixed modes shift in the opposite directions and the degenerate spectra split, that is, the Stark effect.
	%Turning on the external electric field, we find that the meson spectra split, that is, we observe the Stark effect. Note that the fluctuations of one scalar and one component of the gauge field along the electric field are coupled, while the other modes are decoupled. 
	Furthermore, close to the critical electric field one of the mixed modes becomes tachyonic, i.e., a mode with a pure imaginary frequency ($\omega^2<0$).
	This indicates that the background configuration becomes dynamically unstable and the onset of the instability coincides with a turning point of the $E$-$J$ and $E$-$c$ curves. Such behavior repeatedly appears as the electric field is close to the critical value. Although we found similar unstable modes in the dissociation phase in \cite{Ishigaki:2021vyv}, we can observe infinite unstable modes in the meson phase. 
	In each phase near the critical value, the brane embedding exhibits the discrete self-similarity with each period.
	This ensures that emergence of the unstable modes will infinitely repeat toward the critical point.
	%Moreover, we observe an infinite number of purely imaginary modes close to the critical electric field, with half of them being dynamically unstable. It should be noted that, in this paper, we find infinite unstable modes in the meson phase for the first time, although we found similar unstable modes in the meson dissociation phase in \cite{Ishigaki:2021vyv}. 
	On the other hand, the other of the mixed modes as well as decoupled modes never becomes tachyonic but has finite real frequencies even at the critical electric field. These spectra continuously connect between the meson/dissociation phases.
	%On the other hand, if we focus on the modes approaching a finite frequency at the critical electric field, their spectra continuously connect between the meson/dissociation phases.

	Once turning on electric fields, fluctuations on the D$7$-brane cannot be decoupled, so that it is difficult to identify which mesonic excitation each mode in the spectra belongs to, generally.
	However, since every fluctuation is decoupled in both limits of zero electric field and large electric field, we can explore which modes originate from the spectra in the decoupled cases as shown in figures \ref{fig:spectra} and \ref{fig:spectra2}.
	%Due to the coupling between the fluctuations of scalar and gauge field, we generally do not know from which the spectra of them originate.
	%However, since those two fluctuations are decoupled in the limit of large electric field limit, equivalently massless limit, we can explore which modes are connected from those spectra as shown in figure \ref{fig:spectra2}.
	
	%Close to the critical electric field in the meson phase, we observe the avoided crossing of the spectra between neighboring levels as shown in figure \ref{fig:spectra}.
	%It implies that the coupling between those two modes are essential and could not be decoupled by the linear combination. 
	%To further investigate the avoided crossing, we compute the residues of the matrix of the corresponding retarded Green's function in the dual field theory and  find that the only one component is finite at the center of the avoided crossing.
	
	Close to the critical electric field in the meson phase, we observe the avoided crossing of the spectra between neighboring levels as shown in figure \ref{fig:spectra}.%
	\footnote{
		An avoided crossing was also observed in the study of the holographic mesino spectrum with the double-trace deformations \cite{Abt:2019tas}.
		(See, e.g., figure 3 in \cite{Abt:2019tas}.)
		The authors of ref.~\cite{Nakas:2020hyo} discussed on the avoided crossing of ref.~\cite{Abt:2019tas} in an analytical way.
	}
	A similar avoided crossing for the Stark effect had been observed in \cite{PhysRevA.20.2251}, in the 1970s, by using the quantum-mechanical calculation.
	A more recent theoretical and experimental study of the Stark effect can be found in, e.g., \cite{Grimmel_2015}.
	In a system without dissipation, the avoided crossing is a universal phenomenon of spectra depending on a parameter ($E$ in this case).
	%in the adiabatic limit.
	The avoided crossing between two states happens only when these states interact with each other, i.e., their Hamiltonian has non-diagonal components.
	Thus, the avoided crossing in our results implies that the coupling between those two families of modes $n_{\pm}$ are essential, and the corresponding bulk perturbations, $w(t,u)$ and $a_{\parallel}(t,u)$, could not be decoupled by taking any linear combinations for general finite $E$.
	%** duplicate **
	%In our computation, we found that there is a relation between the center of the avoided crossing and the residue matrices.
	%It will be interesting to check the same behavior of the residue matrices at the center of the avoided crossing in the quantum mechanics, and investigate the background mechanism of this relation.
	
	Investigating the residues of the retarded Green's function around the avoided crossing, we find the point $E=\Eac$ at which the residue matrices have only one component while the others vanish (figure \ref{fig:residue2}).
	%As it is shown in figure \ref{fig:residue2}, we find the point $E=\Eac$ where the residue matrices have only one component, and the others vanish.
	We also find that the point $E=\Eac$ almost agrees with the center of the avoided crossing $E=\Enear$.
	One may naively think that such a behavior of the residue matrices indicates decoupling of the modes but it is not true.
	Even at $E=\Eac$, the perturbations, $w(t,u)$ and $a_{\parallel}(t,u)$, are coupled in the bulk as we discussed above.
	Our observations imply that the characteristic behavior of the residue matrices may play a crucial role in determining the center of the avoided crossing.
	%Even for these modes at $E=\Eac$, the perturbations, $w(t,u)$ and $a_{\parallel}(t,u)$, are coupled in the bulk, and the avoided crossing implies that the coupling of the modes are essential as we discussed above.
	%From these observations, we can conclude that the behavior of the residue matrices observed here does not mean the decoupling of the modes.
	%We can also understand the result of  (\ref{eq:residues_at_avoided_crossing})  as follows.
	%If the avoided crossing does not happen, $\omega_{0_{+}}$ can be equal to $\omega_{1_{-}}$.
	%Then, the two contributions in (\ref{eq:residues_at_avoided_crossing}) become one contribution with a regular-diagonal matrix as same as in the decoupled case (\ref{eq:residues_decoupling}).
	%These contributions, however, cannot become one diagonal contribution due to the avoided crossing actually.

\begin{figure}[tbp]
		\centering
		\includegraphics[width=10cm]{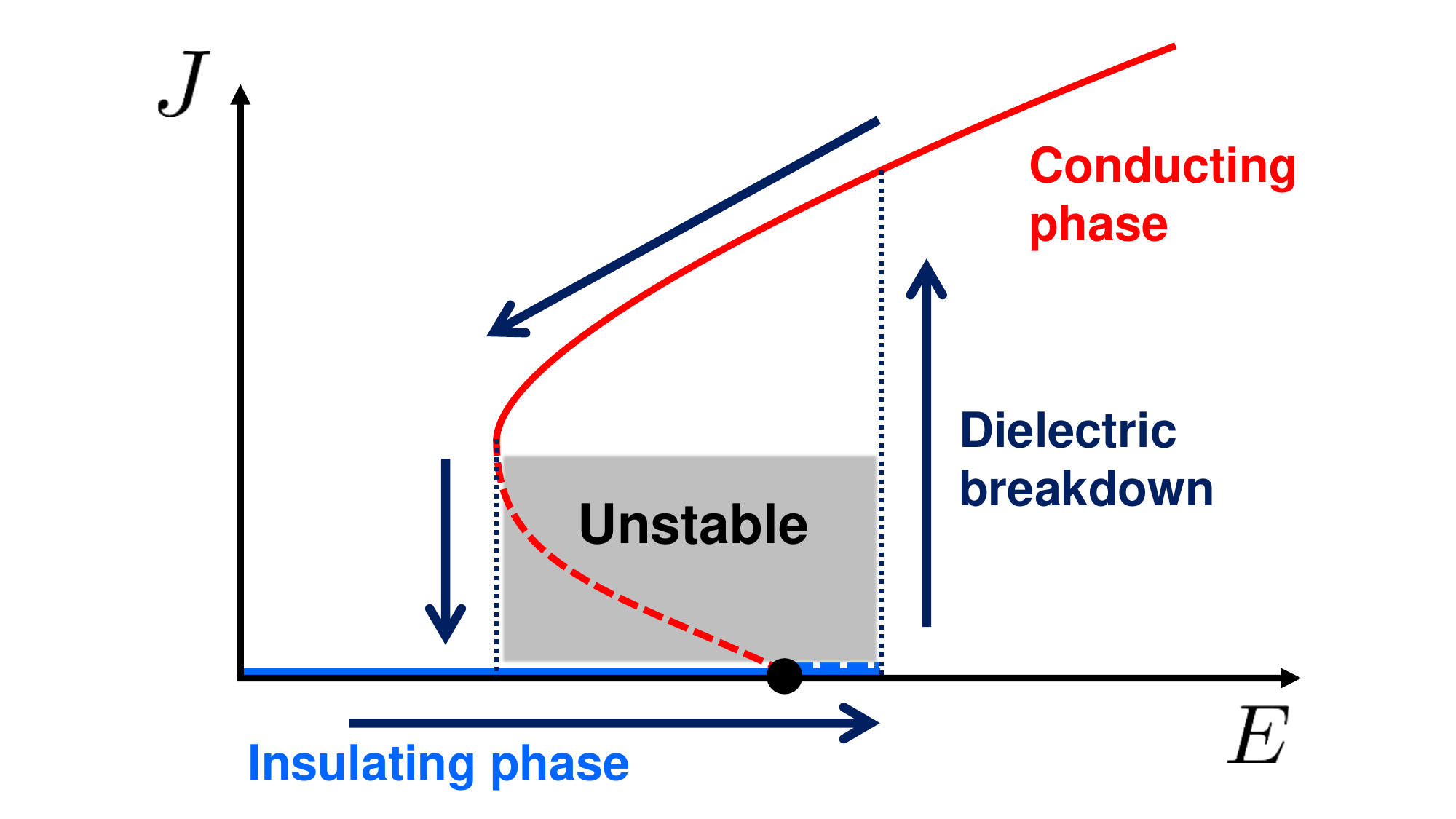}
		\caption{The schematic picture of the hysteresis loop in the $E$-$J$ curve.
		The middle region denoted by the dashed line is dynamically unstable.}
		\label{fig:hysteresis}	
	\end{figure}

	We have revealed dynamical stability over all the values of $E$ at zero temperature, regardless of whether the background state is the meson phase or the dissociation phase. It turns out that each phase remains dynamically stable without any unstable modes until the turning point first emerges on the $E$-$J$ and $E$-$c$ curves.
	It is associated with phenomenon in condensed matter systems. As mentioned in introduction, the meson/dissociation phases correspond to the insulating/conducting phases in the context of condensed matter physics. The meson spectra studied in this paper corresponds to the excitons spectra for finite electric field. Additionally, the instability near the critical electric field implies that there is the hysteresis loop in the $E$-$J$ curve as shown in figure \ref{fig:hysteresis}. The similar hysteresis loop in the $E$-$J$ curve was experimentally observed in an organic compound \cite{Sawano2005organic}.

	We conclude the paper with some remarks on the further investigations.
	In this paper, we focused only on the case of zero R-charge which corresponds to zero angular momentum in the $S^{3}$ \cite{Kruczenski:2003be}.
	By taking it into account, more systematic discussion of the meson spectra at finite electric field would be possible.
	Additionally, it would be straightforward to explore the meson spectra at finite temperature.
	For instance, it would be interesting to study the effect of temperature on the Stark effect.
	Another important direction is to investigate the relationship between an avoided crossing and a residue of the retarded Green's function.
	It would be nice if we could gain a deeper understanding with a simpler coupled system, but it leaves our future work.

\section*{Acknowledgments}
The authors are grateful to Yongjun Ahn and Xin-Meng Wu for comments and discussions.
We also thank Johanna Erdmenger for a fruitful discussion about avoided crossing. 
S.~I. is supported by National Natural Science Foundation of China with Grant No.~12147158.
S.~K. is supported by JSPS Grant-in-Aid for Scientific Research Number~21H05186.
M.~M. is supported by National Natural Science Foundation of China with Grant No.~12047538.

\appendix
\section{The effective action of the perturbation fields}\label{appendix:eom_pert}
In the previous paper \cite{Ishigaki:2020coe}, we showed the quadratic action of the coupled sector for the following perturbations:
\begin{equation}
	\theta \to \theta(u) + \epsilon\vartheta(t,u),\quad
	A_{a}\dd{\xi^{a}} \to (- E t + h(u)) \dd{x} + \epsilon a_{\parallel}(t,u)\dd{x}.
\end{equation}
The quadratic action is given by
\begin{equation}
	S^{(2)} = \mathcal{N} \int \dd[4]{x}\dd{u} \Big[
		\partial_{\alpha} \Psi^T \mathcal{A}^{\alpha\beta} \partial_{\beta} \Psi
  		+ \Psi^T \mathcal{B}^{\alpha} \partial_{\alpha} \Psi
  		+ \Psi^T \mathcal{C} \Psi
	\Big],
	\label{eq:effective_action_Appendix}
\end{equation}
where $\Psi = (\vartheta, a_{\parallel})^{T}$, and
\begin{subequations}
\begin{gather}
	\mathcal{A}^{\alpha\beta}
	=
	\frac{\mathcal{L}}{2}
	\left[
		\gamma^{\alpha\beta}P(u) +
		2 \delta^{[\alpha}_{t} \delta^{\beta]}_u Q(u)
	\right],\\
	\mathcal{B}^{\alpha}
	= 3 \mathcal{L} \tan\theta
	\begin{pmatrix}
		- M^{\alpha u} \theta'(u) & M^{\alpha x}\\
		0 & 0
	\end{pmatrix},\quad
	\mathcal{C} = 
	- \frac{3}{2} \mathcal{L}
	(1 - 2\tan \theta)
	\begin{pmatrix}
		1 & 0\\
		0 & 0
	\end{pmatrix},
\end{gather}
\label{eq:coeffs_position}
\end{subequations}
and
\begin{subequations}
\begin{gather}
	P(u) \equiv
	\begin{pmatrix}
		\Xi(u) & -M^{xu} \theta'(u)\\
		-M^{xu} \theta'(u) & \gamma^{xx}
	\end{pmatrix},\\
	Q(u) \equiv
	(M^{tx} \gamma^{uu} + \gamma^{tu} M^{xu})\theta'(u)
	\begin{pmatrix}
		0 & 1\\
		-1 & 0
	\end{pmatrix},
\end{gather}
\end{subequations}
and $\Xi(u) = 1- \gamma^{uu}\theta'(u)^2$.

In this paper we consider the perturbation fields of (\ref{eq:perturbation_ansatz}).
These are related by
\begin{equation}
	\Psi = \varrho \Phi, \quad
	\varrho \equiv \text{diag}(u, 1).
\end{equation}
The coefficient matrices are related by
\begin{subequations}
\begin{alignat}{3}
	&A^{\alpha\beta}& =& \varrho^T \mathcal{A}^{\alpha\beta} \varrho,\\
	&B^{\alpha}& =& \varrho^T \mathcal{B}^{\alpha} \varrho
	+ 2 (\partial_{\beta} \varphi^T) \mathcal{A}^{\beta\alpha} \varphi,\\
	&C& =& \varrho^T \mathcal{C} \varrho
	+ \varphi^T \mathcal{B}^{\alpha} \partial_{\alpha} \varphi
	+ \partial_{\alpha} \varrho^T \mathcal{A}^{\alpha\beta} \partial_{\beta} \varrho.
\end{alignat}
\end{subequations}
We used $\mathcal{A}^{\alpha\beta} = (\mathcal{A}^{\beta\alpha})^T$.
The coefficient matrices in (\ref{eq:quadratic_action_momentum}) are also related by
\begin{subequations}
\begin{alignat}{3}
	&\bar{A}& =& A^{uu},\\
	&\bar{B}_{\omega}& =& 
	B^{u} - 2 i \omega A^{ut},\\
	&\bar{C}_{\omega}& =& 
	C - i\omega B^{t} +\omega^2 A^{tt}.
\end{alignat}
\end{subequations}
The coefficient matrix of the couterterms is also given by
\begin{equation}
	\bar{C}_{\omega,\epsilon} =
	\frac{\omega^2}{4}
	\begin{pmatrix}
		1 & 0\\
		0 & 1
	\end{pmatrix}
	\log m^2 \epsilon^2.
\end{equation}
Note that since it has a logarithmic dependence on $\epsilon$, the counterterm and the resulting Green's function has ambiguity by shifting of $\log m^2 \epsilon^2 \to \log \Lambda^2 \epsilon^2$ with an arbitrary scale $\Lambda$.
Nevertheless, this ambiguity will not affect the result of calculating the meson spectra. Such a logarithmic contribution originates from the asymptotic behavior of the non-normalizable modes, whereas the normalizable modes corresponding to the mesonic excitations are irrelevant to this term. 

\section{The collocation method}\label{appendix:collocation}
In section \ref{sec:Results}, we use the collocation method, also known as the pseudospectral method, to find (quasi)normal modes numerically.
The collocation method allows us converting a two-points boundary problem of linear differential equations to a simple eigenvalue problem of a large matrix.
We briefly describe this method in this appendix to make our paper self-contained.
One can find further information about this method in ref.~\cite{boyd2001chebyshev} which is known as a standard reference.

\newcommand{\uu}{\mathfrak{u}}
First of all, we introduce a scaled coordinate $\uu \equiv u/\uIR$ where $\uIR = u_*$ or $u_0$ corresponding to whether the background solution is BH or Minkowski embedding, respectively.
The integration region becomes $0\leq \uu\leq 1$, and the boundary conditions for the perturbation fields are the regular condition at $\uu=1$ and the vanishing Dirichlet condition at $\uu=0$ in both the BH and Minkowski embeddings.

Let us consider to approximate the variable $\phi(\uu)$ by the basis function expansion with the Chebyshev polynomial of the first kind $T_{k}(x)$:
\begin{equation}
	\phi(\uu) \approx \sum_{j=0}^{n} \hat{\phi}_j T_{j}(2 \uu - 1),
\end{equation}
where $\hat{\phi}_j$ are spectral coefficients, and $n$ is a finite number of the truncation.
Since $T_{k}(x)$ is regular at $x=-1$ and $1$, we transformed the parameter by $x=2 \uu -1$.
In the collocation method, the variables are discretized at points of a nonuniform grid corresponding to the basis function of the expansion.
In this study, we employ the Gauss-Lobatto-Chebyshev (extrema) grid for $\uu$ spanned in $[0, 1]$ which is given by
\begin{equation}
	\uu_k = \frac{1}{2}\left(
		1 - \cos \frac{k}{n}\pi
	\right),\quad k = 0, 1,\ldots ,n.
\end{equation}
At the collocation points, the expansion will be exact relation:
\begin{equation}
	\phi(\uu_k) = \sum_{j=0}^{n} \hat{\phi}_j T_{j}(2 \uu_k - 1) \equiv \phi_{k}.
\end{equation}
Furthermore, the derivative of $\phi(\uu)$ with respect to $\uu$ on the collocation grid can be performed by multiplying the following differential matrix:%
\footnote{
	Note that the factor of $ -2$ is just a Jacobian arising from changing the parameter from $x$ to $\uu$.
}
\begin{equation}
	\mathcal{D}_{ij} =  -2\times
	\begin{cases}
		\frac{2n^2 + 1}{6} &i=j=0~\text{or}~i=j=n\\
		- \frac{\uu_i}{2(1-\uu_i^2)} &i=j \neq 0,n\\
		\frac{c_i}{c_j}\frac{(-1)^{i+j}}{\uu_i-\uu_j} & \text{otherwise}
	\end{cases},
	\quad
	c_i = \begin{cases}
		2 & k=0, n\\
		1 & \text{otherwise}
	\end{cases}.
\end{equation}
Namely, $\phi'(u_k)$ can be written as $\phi'(u_k) = \sum_{j=0}^n\mathcal{D}_{k j} \phi_j$ by using the differential matrix.
The second derivative can be also performed by multiplying $\mathcal{D}_{ij}$ twice.
By using the differential matrix, we can rewrite the equations of motion at the collocation points.
The regular condition is imposed by interpolating the variables with the Chebyshev polynomials, implicitly.
The vanishing Dirichlet condition can be implemented by replacing the component of the discretized equation by $\phi(\uu_0) = 0$.%
\footnote{
	A more efficient way is writing original equations in terms of $\bar{\phi} = \phi/u$, then the vanishing Dirichlet condition of $\phi$ becomes the regular condition of $\bar{\phi}$ at $u=0$.
	The regular conditions at both ends of the integral region are imposed automatically, hence we do not need to replace the component of the discretized equations.
}
Then, the problem is encoded to solving the algebraic equations for $\phi(\uu_k)$.
When the system is linear, the algebraic equations are also linear.

In practice, we need to perform a few transformations to apply this method to our problem, as follows.
These procedures were explained in \cite{Jansen:2017oag}.
First, we need to deal with ODEs to study the coupled sector.
We formally write the discretized equations of motion for $\phi^{I} = w, a_{\parallel}$ as
\begin{equation}
	0 = \sum_{J = 1}^{2} \sum_{j=0}^{n} \mathsf{L}_{ij}^{IJ} \phi^{I}_{j},
\end{equation}
where indices $I,J = 1,2$ and $i, j = 0,1, \ldots ,n$.
We consider the explicit boundary conditions are already implemented as corresponding rows of $\mathsf{L}_{ij}^{IJ}$.
The equations can be flatten as
\begin{equation}
	0 =
	\begin{pmatrix}
		\mathsf{L}^{11} & \mathsf{L}^{12}\\
		\mathsf{L}^{21} & \mathsf{L}^{22}
	\end{pmatrix}
	\begin{pmatrix}
		\phi^{1} \\ \phi^{2}
	\end{pmatrix}
	\equiv \mathsf{L} \vec{\phi}.
\end{equation}
Here, we suppressed the lower indices, and redefined $2(n+1) \times 2(n+1)$ matrix $\mathsf{L}$ and $2(n+1)$-vector $\vec{\phi}$.
In our model, $\mathsf{L}$ can be decomposed as
\begin{equation}
	\mathsf{L} = \tilde{\mathsf{M}}_{0} + \tilde{\mathsf{M}}_{1} \omega + \tilde{\mathsf{M}}_{2} \omega^2.
\end{equation}
According to \cite{Jansen:2017oag}, we define
\begin{equation}
	\mathsf{M}_{0}
	=
	\begin{pmatrix}
		\tilde{\mathsf{M}}_0 & \tilde{\mathsf{M}}_1\\
		0 & \mathbf{1}
	\end{pmatrix},\quad
	\mathsf{M}_{1}
	=
	\begin{pmatrix}
		0 & \tilde{\mathsf{M}}_{2}\\
		- \mathbf{1} & 0
	\end{pmatrix},
\end{equation}
where $\mathbf{1}$ denotes $2(n+1)\times 2(n+1)$ identity matrix.
The following eigenvalue problem is, then, equivalent to the original problem.%
\footnote{
	If the equation does not involve $\omega$ but only $\omega^2$, we do not need this procedure by regarding $\omega^2$ as an eigenvalue.
	This case corresponds to the Minkowski embedding.
}
\begin{equation}
	\left(\mathsf{M}_{0} + \omega \mathsf{M}_{1}\right) \mathsf{v} = 0.
	\label{eq:generalized_eigenvalue_prob}
\end{equation}
The $4(n+1)$-components eigenvector $\mathsf{v}$ is related with $\vec{\phi}$ by $\mathsf{v} = (\vec{\phi}, \omega \vec{\phi})^{T}$.
The form of (\ref{eq:generalized_eigenvalue_prob}) is called generalized eigenvalue problem.
One can solve such a problem, for example, by using \texttt{Eigenvalues} or \texttt{Eigensystem} built-in function in Mathematica efficiently.

To ensure enough numerical accuracy, we use background solutions which are obtained by the collocation method with the same grid, in the analysis of the perturbation fields.
Since the equation of the background field is nonlinear, the discretized equations are also nonlinear algebraic equations.
The discretized equations can be solved by iterative methods with a suitable initial guess.
In Mathematica, they can be solved with \texttt{FindRoot}, simply.

In this study, we take the number of the collocation points $n=50$ basically.
We have checked that the results are robust even if we vary $n$ in most cases.
Near the critical embedding, however, we need a larger number of the grid for reasonable results.
For obtaining the result to the third turning point in the right panel of figure \ref{fig:avoided_crossing} correctly, we take $n=100$.
It may be caused by that the very small difference of the background solution affect to the result of the perturbations in this regime.

\section{The Wess-Zumino term}\label{appendix:WZ_term}
In this section, we explicitly show that the Wess-Zumino (WZ) term does not give any contribution both in the background and linear perturbation analysis, in our setup.
The WZ term is given by
\begin{equation}
	S_{\text{WZ}}
	= \frac{T_{\text{D}7}}{2}
	\int \mathcal{P}[C_{(4)}] \wedge F \wedge F,
	\label{eq:WZ-term}
\end{equation}
where $\mathcal{P}[C_{(4)}]$ is the pullback of the 4-form Ramond-Ramond potential.
We set $2\pi\alpha'=1$ in this section for simplicity.
The pullback potential is given by
\begin{equation}
	\mathcal{P}[C_{(4)}]
	= u^{-4} \dd{t}\wedge\dd{x}\wedge \dd{y} \wedge \dd{z}
	+ \cos^4\theta \partial_{a} \psi \dd{x^a}\wedge %\mathrm{vol}_{S^3},
	\dd\Omega_3
	,
\end{equation}
where %$\mathrm{vol}_{S^3}$ 
$\dd\Omega_3$
is the volume form of the unit 3-sphere.
Since we do not consider the $S^3$ components of the gauge fields, the WZ term reduces to
\begin{equation}
	S_{\text{WZ}} = \frac{\mathcal{N}}{2} \int \cos^4\theta \partial_{\mu}\psi \dd{x^{\mu}} \wedge F \wedge F,
	\label{eq:WZ-term_5d}
\end{equation}
after integrating the $S^3$ part.
%\begin{equation}
%	F = \frac{F_{tx}}{2} \dd{t}\wedge\dd{x} + \frac{F_{xu}}{2} \dd{x}\wedge\dd{u}.
	%\label{eq:F_components}
%\end{equation}
Our ansatz of the background embedding functions and gauge fields given by eq.~(\ref{eq:ansatz_background}) yields 
\begin{equation}
 \theta = \theta (u), \quad \psi =\text{const.} ,\quad 
	F = \frac{F_{tx}}{2} \dd{t}\wedge\dd{x} + \frac{F_{xu}}{2} \dd{x}\wedge\dd{u}.
	\label{eq:F_components}
\end{equation}

Substituting these expressions into the equations of motion derived from eq.~(\ref{eq:WZ-term_5d}), they vanish.

Next, we consider the linear perturbations for the fields.
Corresponding to eq.~(\ref{eq:perturbation_ansatz}), we write
\begin{equation}
	\theta \to \theta(u) + \epsilon \vartheta(t,u),\quad
	\psi \to \epsilon \psi(t,u),\quad
	A_{\mu} \to A_{\mu}(t,u) + \epsilon a_{\mu}(t,u),
\end{equation}
and $f \equiv \dd{a}$.
The quadratic part of the WZ term is obtained as
\begin{equation}
	S_{\text{WZ}}^{(2)} =
	\frac{\mathcal{N}}{2} \int\left[
		- 4 \cos^3\theta\sin\theta ~ \vartheta ~ \partial_{\mu}\psi \dd{x^{\mu}} \wedge F \wedge F
		+ 2 \cos^4\theta \partial_{\mu}\psi \dd{x^{\mu}}\wedge F\wedge f
	\right].
	\label{eq:quadratic_WZ}
\end{equation}
Using eq.~(\ref{eq:F_components}), the first term immediately vanishes because $F\wedge F = 0$.
Taking the coordinate dependence of the perturbations into account, the second term of the integrand can be expressed as
\begin{equation}
	\cos^4\theta \left(
		\partial_{t} \psi \times f_{u \mu} \dd{t} \wedge F \wedge \dd{u}
		+ \partial_{u} \psi \times f_{t \mu} \dd{u} \wedge F \wedge \dd{t}
	\right)\wedge \dd{x^{\mu}}.
\end{equation}
This also vanishes with eq.~(\ref{eq:F_components}).
Therefore, the WZ term has no contribution for the analysis of the linear perturbations in our setup.
In order for the second term in eq.~(\ref{eq:quadratic_WZ}) to survive, the term $\dd\psi\wedge f$ should have $tyz$- or $uyz$-components, for example. This implies that the WZ term may contribute when perturbations are spatially inhomogeneous.

\bibliography{main}
\bibliographystyle{JHEP}
\end{document}